\documentclass[a4,12pt]{article}
\usepackage[dvips]{graphicx,psfrag}
\usepackage[abs]{overpic}
\usepackage{float}
\def\dags{{\protect\mbox{\tiny \dag}}}

\def\ls{\lower0.5ex\hbox{$\buildrel >\over{\scriptstyle\sim}$}} 
\def\rs{\lower0.5ex\hbox{$\buildrel <\over{\scriptstyle\sim}$}} 
\begin{document}
\pagestyle{empty} \setlength{\footskip}{2.0cm}
\setlength{\oddsidemargin}{0.5cm}
\setlength{\evensidemargin}{0.5cm}
\renewcommand{\thepage}{-- \arabic{page} --}
\def\mib#1{\mbox{\boldmath $#1$}}
\def\bra#1{\langle #1 |}  \def\ket#1{|#1\rangle}
\def\vev#1{\langle #1\rangle} \def\dps{\displaystyle}
\newcommand{\fcal}{{\cal F}}
\newcommand{\gcal}{{\cal G}}
\newcommand{\ocal}{{\cal O}}
\newcommand{\El}{E_\ell}
\renewcommand{\thefootnote}{$\sharp$\arabic{footnote}}
\newcommand{\W}{{\scriptstyle W}}
 \newcommand{\I}{{\scriptscriptstyle I}}
 \newcommand{\J}{{\scriptscriptstyle J}}
 \newcommand{\K}{{\scriptscriptstyle K}}
%
% ------------------------------------------------------------
 \def\thebibliography#1{\centerline{REFERENCES}
 \list{[\arabic{enumi}]}{\settowidth\labelwidth{[#1]}\leftmargin
 \labelwidth\advance\leftmargin\labelsep\usecounter{enumi}}
 \def\newblock{\hskip .11em plus .33em minus -.07em}\sloppy
 \clubpenalty4000\widowpenalty4000\sfcode`\.=1000\relax}\let
 \endthebibliography=\endlist
 \def\sec#1{\addtocounter{section}{1}\section*{\hspace*{-0.72cm}
 \normalsize\bf\arabic{section}.$\;$#1}\vspace*{-0.3cm}}
 \def\subsec#1{\addtocounter{subsection}{1}\subsection*{\hspace*{-0.4cm}
 \normalsize\bf\arabic{section}.\arabic{subsection}.$\;$#1}\vspace*{-0.3cm}}
% ------------------------------------------------------------
% \vspace*{-1.7cm}
% \noindent
% \phantom
% {\large\bf July 16, 2007}

\vspace*{-1.7cm}
\begin{flushright}
$\vcenter{
%%% \phantom{  \hbox{{\footnotesize FUT and TOKUSHIMA Report}}  }
\hbox{{\footnotesize FUT and TOKUSHIMA Report}}
%%% \phantom{  {\hbox{{\footnotesize TOKUSHIMA Report}}}  }
{  \hbox{(arXiv:1104.1221)}  }
}$
\end{flushright}

\vskip 0.8cm
\begin{center}
\hspace*{-0.6cm}
$\mbox{\large\bf Exploring anomalous top-quark interactions via the final lepton}$

\vskip 0.18cm
{\large\bf in ${\mib t}\bar{{\mib t}}$ productions/decays at hadron colliders}

\end{center}

\vspace{0.9cm}
\begin{center}
\renewcommand{\thefootnote}{\alph{footnote})}
Zenr\=o HIOKI$^{\:1),\:}$\footnote{E-mail address:
\tt hioki@ias.tokushima-u.ac.jp}\ and\
Kazumasa OHKUMA$^{\:2),\:}$\footnote{E-mail address:
\tt ohkuma@fukui-ut.ac.jp}
\end{center}

\vspace*{0.4cm}
\centerline{\sl $1)$ Institute of Theoretical Physics,\
University of Tokushima}

\centerline{\sl Tokushima 770-8502, Japan}

\vskip 0.2cm
\centerline{\sl $2)$ Department of Information Science,\
Fukui University of Technology}
\centerline{\sl Fukui 910-8505, Japan}

\vspace*{2.3cm}
\centerline{ABSTRACT}

\vspace*{0.2cm}
\baselineskip=21pt plus 0.1pt minus 0.1pt
We study momentum distributions of the final-state charged lepton in $p\bar{p}/pp$ 
$\to t\bar{t} \to \ell^+ X\:(\ell=e\ {\rm or}\ \mu)$ at hadron colliders, i.e., Tevatron
and Large Hadron Collider (LHC) in order to explore possible new-physics effects in the
top-quark sector. Assuming general model-independent $t\bar{t}g+t\bar{t}gg$ and $tbW$
interactions beyond the standard model, we first derive analytical formulas for the
corresponding parton-parton processes. We then compute the lepton angular, energy and
transverse-momentum distributions in $p\bar{p}/pp$ collisions to clarify how they are
affected by those anomalous couplings.

\vskip 1.5cm
\centerline{( Published in {\it Phys. Rev.} {\bf D83} (2011), 114045. )}

\vfill
% PACS:  12.38.-t, 12.38.Bx, 12.38.Qk, 12.60.-i, 14.65.Ha, 14.70.Dj

% Keywords:
% Hadron colliders, Anomalous couplings, Top productions,\\ \ \ \ \ \ \  Lepton distributions

\newpage
\renewcommand{\thefootnote}{$\sharp$\arabic{footnote}}
%-------------------------------------------------------------
\pagestyle{plain} \setcounter{footnote}{0}

% 111111111111111111111111111111111111111111111111111111111111
\sec{Introduction}

The Large Hadron Collider, LHC, has started to operate and is already giving us new data
\cite{LHC}. Thereby we will soon be able to explore physics beyond the standard model of
the strong and electroweak interactions in TeV world with high precision. Studies of such
new physics can be classified into two categories: model-dependent and model-independent
approaches. The former approach could enable precise calculations but require us to start
everything from the beginning if the wrong model was chosen. In contrast to it, we would
rarely fail to get meaningful information in the latter but it would not be that easy
there to perform precision analyses since we usually need to treat many unknown parameters
together.
 
One reasonable way to decrease the number of such unknown parameters in a model-independent
analysis is to assume a new physics characterized by an energy scale ${\mit\Lambda}$
and write down $SU(3)\times SU(2)\times U(1)$-symmetric effective operators for the world
below ${\mit\Lambda}$. Those operators with dimension 6, having the leading importance,
were listed in \cite{Buchmuller:1985jz},
supposing there exist only standard particles below ${\mit\Lambda}$. Although we still have
to treat several operators (parameters) even in this framework, but some of the operators
given there were found to be dependent on each other through equations of motion
\cite{Grzadkowski:2003tf}. This shows that we might further be able to reduce the number of
independent operators, and indeed it was done in \cite{AguilarSaavedra:2008zc} (see also
\cite{Grzadkowski:2010es}).

In this effective-operator framework, not only electroweak couplings but also QCD couplings
receive nonstandard corrections. Although it might be hard to imagine that the QCD
couplings of light quarks are affected by those anomalous interactions, 
the top-quark couplings could however be exceptional, because this quark has not been studied
enough precisely yet, and its extremely heavy mass seems to tell us something about a new
physics beyond the standard model. That is, this heaviest quark might be able to work as a
precious window to a nonstandard physics once LHC starts to give us fruitful data.

Under this consideration, we have performed analyses of anomalous top-gluon couplings
produced by the dimension-6 operators through $t\bar{t}$ productions at Tevatron
and LHC \cite{Hioki:2009hm,Hioki:2010zu}. In this article, we would like to develop them and
perform more realistic analyses focusing on momentum distributions of the charged lepton
coming from the semileptonic top decay in $p\bar{p}/pp \to t\bar{t}X$. In fact, a number of
authors have studied top anomalous interactions through such final leptons in $p\bar{p}/pp$
collisions \cite{Brandenburg:1992be}--\cite{Zhang:2010dr}. Their main interests are in
$C\!P$-violation, which is reasonable since the standard-model $C\!P$-violation in the top
sector is expected so small that any non-negligible $C\!P$-violation effects there will be
a signal of new physics.

Here, however, we have no mind to add similar studies to them. Our main purpose of revisiting
this issue is to present analytical formulas of the final-lepton momentum distributions for
practical use, and carry out some computations and analyses based on them, taking into account
both anomalous $C\!P$-conserving {\it and} $C\!P$-violating top-gluon couplings {\it plus}
nonstandard $tbW$ coupling altogether.

We first describe our calculational framework in section 2. In section 3, we derive the
final-lepton momentum distribution in partonic processes $q\bar{q}/gg \to t\bar{t} \to \ell^+ X$,
and present their analytical expressions. We then transform them into the angular, energy and
transverse-momentum distributions of the final charged lepton in the hadronic processes
$p\bar{p}/pp \to t\bar{t}X \to \ell^+ X'$, and study those distributions numerically for several
typical parameter sets in section 4. Finally, a summary and some remarks are given in section 5.
In the appendix, we refine our last result on the anomalous top-gluon couplings \cite{Hioki:2010zu}
with the Tevatron and CMS data \cite{Teva-data,Khachatryan:2010ez} by adding the latest ATLAS data
\cite{Aad:2010ey}.

% 222222222222222222222222222222222222222222222222222222222222
\sec{Framework}

Let us describe our basic framework in this section. In ref.\cite{Buchmuller:1985jz} were given
three effective operators contributing to top-gluon interactions. Those operators produce top-pair 
production amplitudes which include $\gamma^\mu$, $\sigma^{\mu\nu}q_\nu$, $(p_i + p_j)^\mu$
and $q^\mu$ terms (or more complicated Lorentz structure), where $p_{i,j}$ and $q$ are
the top-quark $i$, $j$ and gluon momenta. However two of them were shown not to be
independent in \cite{AguilarSaavedra:2008zc}, and we only need to take into account
one operator
\begin{equation}
{\cal O}^{33}_{uG\phi}
=\sum_a [\:\bar{q}_{L3}(x)\lambda^a \sigma^{\mu\nu} u_{R3}(x)
\tilde{\phi}(x) G^a_{\mu\nu}(x)\:],
\end{equation}
where we followed the notation of \cite{AguilarSaavedra:2008zc}: $q_{L3}$ is the third generation of
left-handed $SU(2)$-doublet, i.e., $(t,b)_L^t$, $u_{R3}$ is the third generation up-type
$SU(2)$ singlet, i.e., $t_R$, $\tilde{\phi}\equiv i\tau^2\phi^*$ with $\phi$ being the Higgs doublet,
$G^a_{\mu\nu}$ is the $SU(3)$ gauge-field (=gluon) tensor.
% \[
% G_{\mu\nu}^a=\partial_\mu G_\nu^a-\partial_\nu G_\mu^a
% -g_s \sum_{b,c} f_{abc}G_\mu^b G_\nu^c,
% \]
% where $g_s$ and $f_{abc}$ are the $SU(3)$ coupling constant and structure constant, respectively.

Now the top-gluon interaction Lagrangian including the above operator is given by
\begin{equation}
{\cal L}
={\cal L}_{\rm SM}
+ \frac1{{\mit\Lambda}^2}
(C^{33}_{uG\phi}{\cal O}^{33}_{uG\phi} + C^{33*}_{uG\phi}{\cal O}^{33\dags}_{uG\phi}),
\end{equation}
where ${\cal L}_{\rm SM}$ on the right-hand side means the standard-model QCD top-gluon couplings
and $C^{33}_{uG\phi}$, the coefficient of ${\cal O}^{33}_{uG\phi}$, represents the contribution of
this operator. In our framework this coefficient (plus its complex conjugate) and ${\mit\Lambda}^{-2}$
are combined and treated as parameters to be determined by experimental data. Since ${\cal O}^{33}_{uG\phi}$
contains $G^a_{\mu\nu}$, the resultant nonstandard interaction has not only $t\bar{t}g$ but also
$t\bar{t}gg$ couplings. Let us therefore denote this Lagrangian by ${\cal L}_{t\bar{t}g,gg}$
hereafter and reexpress it as
\begin{eqnarray}
&&{\cal L}_{t\bar{t}g,gg}
=-\frac12 g_s \sum_a \Bigl[\, \bar{\psi}_t(x) \lambda^a \gamma^\mu \psi_t(x) G^a_\mu(x) \nonumber\\
&&\phantom{{\cal L}_{t\bar{t}g,gg}=-\frac12 g_s \sum_a}
\ \ \ \
- \bar{\psi}_t(x) \lambda^a \frac{\sigma^{\mu\nu}}{m_t} (d_V+id_A \gamma_5) \psi_t(x)
G^a_{\mu\nu}(x)\,\Bigr],  \label{Lag}
\end{eqnarray}
where $g_s$ is the $SU(3)$ coupling constant and $d_{V,A}$ are defined as
\[
d_V \equiv \frac{\sqrt{2}vm_t}{g_s{\mit\Lambda}^2} {\rm Re}(C^{33}_{uG\phi}),\ \ \ \ \
% {\rm and}\ \ \
d_A \equiv \frac{\sqrt{2}vm_t}{g_s{\mit\Lambda}^2} {\rm Im}(C^{33}_{uG\phi})
\]
corresponding to the top chromomagnetic- and chromoelectric-dipole moments respectively
with $v$ being the Higgs vacuum expectation value ($=246$ GeV). Concerning the other light
quarks, i.e., $u$, $d$, $s$, $c$ and $b$, we assume their couplings with the gluon are
properly described by the standard QCD Lagrangian though in
principle there also could be nonstandard corrections in our framework, because those 
couplings have so far been tested very well based on a lot of experimental data.

On the other hand, dimension-6 operators which contribute to top-decay $t\to bW$ are
\begin{eqnarray}
&&{\cal O}^{(3,33)}_{\phi q}
=i\sum_{\I}\,[\,\phi^{\dags}(x)\tau^{\I} D_\mu \phi(x)\,] % \nonumber \\
% &&\ \ \ \ \ \ \ \ \ \ \ \ \ \ \ \
% \times
[\,\bar{q}_{L3}(x)\gamma^\mu \tau^{\I} q_{L3}(x)\,]  \\
&&{\cal O}^{33}_{\phi \phi}
=i[\,\tilde{\phi}^{\dags}(x) D_\mu \phi(x)\,]
[\,\bar{u}_{R3}(x)\gamma^\mu d_{R3}(x)\,]   \\
&&{\cal O}^{33}_{uW}
=\sum_{\I} \,\bar{q}_{L3}(x) \sigma^{\mu\nu} \tau^{\I} u_{R3}(x)
\tilde{\phi}(x) W^{\I}_{\mu\nu}(x)~~~~~~   \\
&&{\cal O}^{33}_{dW}
=\sum_{\I} \:\bar{q}_{L3}(x) \sigma^{\mu\nu} \tau^{\I} d_{R3}(x)
\phi(x) W^{\I}_{\mu\nu}(x),
\end{eqnarray}
where $D_\mu$ is the $SU(2)\times U(1)$ covariant derivative, $d_{R3}$ is the third generation down-type
$SU(2)$ singlet (i.e., $b_R$), and $W^{\I}_{\mu\nu}$ is the $SU(2)$ gauge-field tensor.
% \[
% W_{\mu\nu}^{\I}=\partial_\mu W_\nu^{\I}-\partial_\nu W_\mu^{\I}
% -g \sum_{\J,\K} \epsilon_{\I\!\J\!\K}W_\mu^{\J} W_\nu^{\K}
% \]
% with $g$ and $\epsilon_{\I\!\J\!\K}$ being $SU(2)$ coupling and structure constant.

We thereby have the corresponding interaction Lagrangian
\begin{equation}
{\cal L}
={\cal L}_{\rm SM}
+ \frac1{{\mit\Lambda}^2} \sum_i
(C_i{\cal O}_i + C^*_i{\cal O}^{\dags}_i),
\end{equation}
where ${\cal L}_{\rm SM}$ gives the standard-model $tbW$ couplings this time
and the sum is taken over the above four operators. We denote this Lagrangian by
${\cal L}_{tbW}$ and again reexpress as
\begin{eqnarray}
&&{\cal L}_{tbW} =-{g\over\sqrt{2}}\:\Bigl[\,
\bar{\psi}_b(x)\gamma^{\mu}(f_1^L P_L +f_1^R P_R)\psi_t(x) W_\mu^-(x) \nonumber \\
&&\phantom{{\cal L}_{tbW} =-{g\over\sqrt{2}}}\ \ \
+\,\bar{\psi}_b(x){{\sigma^{\mu\nu}}\over M_W}
(f_2^L P_L +f_2^R P_R)\psi_t(x) \partial_\mu W_\nu^-(x)\,\Bigr],
\label{LagW}
\end{eqnarray}
where $g$ is the $SU(2)$ coupling constant, $P_{L/R}\equiv(1\mp\gamma_5)/2$,
\[
\begin{array}{ll}
f_1^L \equiv V_{tb}+C_{\phi q}^{(3,33)*} \displaystyle{\frac{v^2}{{\mit\Lambda}^2}},
&\ \ \ \
f_1^R \equiv C_{\phi \phi}^{33*} \displaystyle{\frac{v^2}{2{\mit\Lambda}^2}}, \\
% & \\
f_2^L \equiv -\sqrt{2}C_{dW}^{33*} \displaystyle{\frac{v^2}{{\mit\Lambda}^2}},
&\ \ \ \
f_2^R \equiv -\sqrt{2}C_{uW}^{33} \displaystyle{\frac{v^2}{{\mit\Lambda}^2}} \\
\end{array}
\]
and $V_{tb}$ is $(tb)$ element of Kobayashi-Maskawa matrix. 
Here again $\ell\nu W$ couplings, which become necessary for $W^+\to\ell^+\nu_\ell$ occurring
after $t\to bW^+$, could also have nonstandard terms, but we adopt
the SM Lagrangian for this part due to the same reason as the light-quark and gluon
interactions.

In calculating the momentum distributions of the final charged lepton in the partonic level,
we utilize the Kawasaki-Shirafuji-Tsai formalism \cite{Kawasaki:1973hf,Tsai:1971vv}.
This formalism is quite valuable when we study the momentum distribution of a
final-state particle from productions/decays of a heavy particle whose mass $m$ and
total width ${\mit\Gamma}$ satisfy $m\gg{\mit\Gamma}$, and consequently ``narrow-width
approximation''
\[
\left|\,{1\over{p^2-m^2+im{\mit\Gamma}}}\,\right|^2
\simeq{\pi\over{m{\mit\Gamma}}}\delta(p^2 -m^2)
\]
holds as a good approximation. 
In this framework, the final-lepton momentum distribution in a collision of particles $a$
and $b$ like $ab \to t\bar{t} \to \ell^+ X$
is given by
\begin{equation}
\frac{d\sigma}{d^3\mib{p}_\ell}(ab \to t\bar{t} \to \ell^+X)
=4\int d{\mit\Omega}_t
\frac{d\sigma}{d{\mit\Omega}_t}(n,0)\frac{1}{{\mit\Gamma}_t}
\frac{d{\mit\Gamma}_\ell}{d^3\mib{p}_\ell}(t\to b\ell^+\nu),~~
\label{master}
\end{equation}
where ${\mit\Gamma}_\ell$ is the width of an unpolarized top, $d\sigma(n,0)/d{\mit\Omega}_t$ is
that obtained from the $t\bar{t}$-production cross section with spin vectors $s_t$ and $s_{\bar{t}}$,
$d\sigma(s_t,s_{\bar{t}})/d{\mit\Omega}_t$, through the following replacement:
\begin{equation}
\phantom{\Bigl(}
s_t^\mu \to n^\mu = \frac{m_t}{p_t p_\ell} p_\ell^\mu  -  \frac1{m_t} p_t^\mu\,,
\ \ \ \ \ s_{\bar{t}}^\mu \to 0.~~
\phantom{\Bigr)}
\label{replacement}
\end{equation}
We get the $\ell^-$ distribution from $\bar{t}$ decay by exchanging the roles of $s_t$ and $s_{\bar{t}}$,
and reversing the sign of $n$.

% 333333333333333333333333333333333333333333333333333333333333
\sec{Parton-process cross sections}

Let us derive the lepton-momentum distribution in the parton processes $q\bar{q}/gg\to t\bar{t}\to \ell^+X$
using interactions (\ref{Lag}) and (\ref{LagW})
in Kawasaki-Shirafuji-Tsai framework. In \cite{Berdine:2007uv} is pointed out that we have to be
careful in applying the narrow-width approximation to a certain process, but in the case of the top
quark and $W$ boson, necessary conditions are satisfied as
$m_t\,(=172.0\pm 1.6\,{\rm GeV}) \gg {\mit\Gamma}_t\,(=1.99^{+0.69}_{-0.55}\,{\rm GeV})$ and
$M_W\,(=80.399\pm 0.023\,{\rm GeV}) \gg {\mit\Gamma}_W\,(=2.085\pm 0.042$ GeV) \cite{PDG,Abazov:2010tm}.

In the following calculations, we neglect all the fermion masses except the top, and
put $V_{tb}$ to be 1 \cite{Aaltonen:2010jr}. In addition, we take into account all the
contributions from $d_{V,A}$ since this is part of the strong interaction, although LHC data
have narrowed the allowed region for them \cite{Hioki:2010zu}, while we include only linear
terms in anomalous $f_{1,2}^{L,R}$ like in \cite{Antipin:2008zx} considering that this is electroweak
interaction and also that all Tevatron data on $t\to bW$ \cite{Abazov:2008sz,Aaltonen:2010ha} are
consistent with the standard model (i.e., $f_1^L=1,\,f_1^R=f_2^{L,R}=0$).

Under these approximations, the $q\bar{q}/gg \to t\bar{t}$ differential cross sections are
\begin{eqnarray}
&&\frac{d\sigma_{q\bar{q}}}{\!\!\!d{\mit\Omega}_t}(s_t,0)
= \frac{\hat{\beta}\alpha_s^2}{36\hat{s}} \Bigl[\,1-2(v-z)-8(d_V-d_V^2+d_A^2)
+8(d_V^2+d_A^2)v/z\,\Bigr]~~~~ \label{qqtt} \\
&&\frac{d\sigma_{gg}}{\!\!\!d{\mit\Omega}_t}(s_t,0)
= \frac{\hat{\beta}\alpha_s^2}{384\hat{s}}
\Bigl[\,(4/v-9)\,[\,1-2v+4z(1-z/v)-8d_V(1-2d_V)\,] \nonumber\\
&&\ \ \ \ \ \ \ \ \ \ \ \ \ \
+4(d_V^2+d_A^2)\,[\,14(1-4d_V)/z+(1+10d_V)/v\,] \nonumber \\
&&\ \ \ \ \ \ \ \ \ \ \ \ \ \
-32(d_V^2+d_A^2)^2(1/z-1/v-4v/z^2)\,\Bigr]
\label{ggtt}
\end{eqnarray}
in the CM frame, and the $t \to bW \to b\ell^+ \nu$ differential width is
\begin{equation}
\frac{1}{{\mit\Gamma}_t}\frac{d{\mit\Gamma}_\ell}{d^3\mib{p}_\ell}
=\frac{6 B_\ell}{\pi m_t^2 {\W} \El}
\omega\Bigl[\,1+2 d_R \Bigl(\frac{1}{1-\omega}-\frac{3}{1+2r} \Bigr)\,\Bigr],~~~
\label{t-decay}
\end{equation}
where $\El$ is the $\ell^+$ energy, i.e., $p_\ell^\mu=(\El,\,\mib{p}_\ell)$,
\begin{eqnarray*}
&&z \equiv m_t^2/\hat{s},\ \ \ v\equiv (\hat{t}-m_t^2)(\hat{u}-m_t^2)/\hat{s}^2,\ \ \
\omega\equiv (p_t -p_\ell)^2/m_t^2,\\
&&r \equiv (M_W/m_t)^2, \ \ \ {\W} \equiv (1-r)^2(1+2r),\ \ \
d_R \equiv {\rm Re}(f_2^R) \sqrt{r},
\end{eqnarray*}
$\hat{s},\hat{t},\hat{u}$ are the Mandelstam variables, $\hat{\beta} \equiv \sqrt{1-4m_t^2/\hat{s}}$ is
the size of the top velocity, $B_\ell$ is the top-semileptonic-decay branching ratio
($={\mit\Gamma}_\ell/{\mit\Gamma}_t$), and we applied the narrow-width approximation to the $W$ propagator.
Here we attached ``\lower1.1ex\hbox{\Large \verb+^+}'' to some variables to clarify that they are parton-level
ones. Note that neither $d\sigma_{q\bar{q}}$ nor $d\sigma_{gg}$ has $s_t$-dependent terms actually.

Combining all those formulas and quantities with eq.(\ref{master}) we arrive at the lepton-momentum
distributions:
\begin{eqnarray}
&&\frac{\ d\sigma_{ab}}{d\El dc_\ell}
= \Bigl[\frac{\ d\sigma_{ab}}{d\El dc_\ell}\Bigr]_{\rm SM}
+ \Bigl[\frac{\ d{\mit\Delta}\sigma_{ab}}{d\El dc_\ell}\Bigr]_{\rm BSM}
% \\
%&&\frac{\ d\sigma_{gg}}{d\El dc_\ell}
%= \Bigl[\frac{\ d\sigma_{gg}}{d\El dc_\ell}\Bigr]_{\rm SM}
%+ \Bigl[\frac{\ d{\mit\Delta}\sigma_{gg}}{d\El dc_\ell}\Bigr]_{\rm BSM},
\end{eqnarray}
where $ab=q\bar{q}$ or $gg$, the first (SM)/second (BSM) terms on the right-hand side
express respectively the standard-model/beyond-the-standard-model contributions, and
$c_\ell$ is our abbreviation of $\cos\theta_\ell$ with $\theta_\ell$ being
the scattering angle between the momenta of the incident parton $a$ and $\ell^+$
(similarly $s_\ell \equiv \sin\theta_\ell$ hereafter). We give their
analytical forms explicitly in the following:
\begin{eqnarray}
&&\Bigl[\frac{\ d\sigma_{q\bar{q}}}{d\El dc_\ell}\Bigr]_{\rm SM}
=\frac{4\hat{\beta}\alpha_s^2}{3m_t^2 \hat{s}} \frac{B_\ell}{{\W}}
E_\ell \Bigl[\,(1+2z)\fcal_0(\El,c_\ell)-2\fcal_1(\El,c_\ell)\,\Bigr] \\
&&\Bigl[\frac{\ d{\mit\Delta}\sigma_{q\bar{q}}}{d\El dc_\ell}\Bigr]_{\rm BSM}
=\frac{4\hat{\beta}\alpha_s^2}{3m_t^2 \hat{s}} \frac{B_\ell}{{\W}}
E_\ell \biggl[\,2d_R\Bigl[\,(1+2z)
\Bigl(\gcal_0(\El,c_\ell)-\frac3{1+2r}\fcal_0(\El,c_\ell)\Bigr)~~~~~
\nonumber\\
&&\ \ \ \ \ \ \ \ \ \ \ \ \ \ \ \ \ \ \ \ \ \ \ \ \ \ \ \ \ \ \ \
-2\Bigl(\gcal_1(\El,c_\ell)-\frac3{1+2r}\fcal_1(\El,c_\ell)\Bigr)\,\Bigr] \nonumber\\
&&\ \ \ \ \ \ \ \ \ \ \ \ \ \ \ \ \ \ \ \
-8(d_V-d_V^2+d_A^2)\fcal_0(\El,c_\ell)
+\frac8{z}(d_V^2+d_A^2)\fcal_1(\El,c_\ell) \nonumber\\
&&\ \ \ \ \ \ \ \ \ \ \ \ \ \ \ \ \ \ \ \
-16d_R(d_V-d_V^2+d_A^2)
\Bigl(\gcal_0(\El,c_\ell)-\frac3{1+2r}\fcal_0(\El,c_\ell)\Bigr) \nonumber\\
&&\ \ \ \ \ \ \ \ \ \ \ \ \ \ \ \ \ \ \ \
+\frac{16}{z}d_R(d_V^2+d_A^2)
\Bigl(\gcal_1(\El,c_\ell)-\frac3{1+2r}\fcal_1(\El,c_\ell)\Bigr)\,\biggr] \\
&&\Bigl[\frac{\ d\sigma_{gg}}{d\El dc_\ell}\Bigr]_{\rm SM}
= \frac{\hat{\beta}\alpha_s^2}{8m_t^2 \hat{s}} \frac{B_\ell}{{\W}}
E_\ell \Bigl[\,-(17+36z)\fcal_0(\El,c_\ell)+18\fcal_1(\El,c_\ell) \nonumber\\
&&\ \ \ \ \ \ \ \ \ \ \ \ \ \ \ \ \ \ \ \
+4(1+4z+9z^2)\fcal_{-1}(\El,c_\ell)-16z^2\fcal_{-2}(\El,c_\ell) \,\Bigr] \\
&&\Bigl[\frac{\ d{\mit\Delta}\sigma_{gg}}{d\El dc_\ell}\Bigr]_{\rm BSM}
= \frac{\hat{\beta}\alpha_s^2}{8m_t^2 \hat{s}} \frac{B_\ell}{{\W}}
E_\ell \biggl[\,2d_R \Bigl[-(17+36z)  % \nonumber \\
% &&\ \ \ \ \ \ \ \ \ \ \ \ \ \ \ \ \ \ \ \ \ \ \ \ \ \ \ \ \ \ \ \ \ \ \ \ \ \ \ \
% \times
\Bigl(\gcal_0(\El,c_\ell)-\frac3{1+2r}\fcal_0(\El,c_\ell)\Bigr) \nonumber\\
&&\ \ \ \ \ \ \ \ \ \ \ \ \ \ \ \ \ \ \ \ \ \ \ \ \ \ \ \ \ \ \ \
+18 \Bigl(\gcal_1(\El,c_\ell)-\frac3{1+2r}\fcal_1(\El,c_\ell)\Bigr) \nonumber \\
&&\ \ \ \ \ \ \ \ \ \ \ \ \ \ \ \ \ \ \ \ \ \ \ \ \ \ \ \ \ \ \ \
+4(1+4z+9z^2)
\Bigl(\gcal_{-1}(\El,c_\ell)-\frac3{1+2r}\fcal_{-1}(\El,c_\ell)\Bigr) \nonumber\\
&&\ \ \ \ \ \ \ \ \ \ \ \ \ \ \ \ \ \ \ \ \ \ \ \ \ \ \ \ \ \ \ \
-16z^2
\Bigl(\gcal_{-2}(\El,c_\ell)-\frac3{1+2r}\fcal_{-2}(\El,c_\ell)\Bigr)\,\Bigr] \nonumber\\
&&\ \ \ \ \ \ \ \ \ \ \ \ \ \ \ \ \ \ \ \
-8d_V(1-2d_V)\Bigl(4\fcal_{-1}(\El,c_\ell)-9\fcal_0(\El,c_\ell)\Bigr) \nonumber \\
&&\ \ \ \ \ \ \ \ \ \ \ \ \ \ \ \ \ \ \ \
+4(d_V^2+d_A^2)
\Bigl(\frac{14}{z}(1-4d_V)\fcal_0(\El,c_\ell)+(1+10d_V)\fcal_{-1}(\El,c_\ell)\Bigr) \nonumber \\
&&\ \ \ \ \ \ \ \ \ \ \ \ \ \ \ \ \ \ \ \
-32(d_V^2+d_A^2)^2
\Bigl(\frac1{z}\fcal_0(\El,c_\ell)-\fcal_{-1}(\El,c_\ell)-\frac4{z^2}\fcal_1(\El,c_\ell)\Bigr)
\nonumber \\
&&\ \ \ \ \ \ \ \ \ \ \ \ \ \ \ \ \ \ \ \
-16d_R d_V(1-2d_V)
\Bigl[\,
4\Bigl(\gcal_{-1}(\El,c_\ell)-\frac3{1+2r}\fcal_{-1}(\El,c_\ell)\Bigr) \nonumber \\
&&\ \ \ \ \ \ \ \ \ \ \ \ \ \ \ \ \ \ \ \ \ \ \ \ \ \ \ \ \ \ \ \
-9\Bigl(\gcal_0(\El,c_\ell)-\frac3{1+2r}\fcal_0(\El,c_\ell)\Bigr)
\,\Bigr] \nonumber \\
&&\ \ \ \ \ \ \ \ \ \ \ \ \ \ \ \ \ \ \ \
+8d_R(d_V^2+d_A^2) % \nonumber \\
% &&\ \ \ \ \ \ \ \ \ \ \ \ \ \ \ \ \ \ \ \ \ \ \ \ \ \ \ \ \ \ \ \
% \times
\Bigl[\,\frac{14}{z}(1-4d_V)
\Bigl(\gcal_0(\El,c_\ell)-\frac3{1+2r}\fcal_0(\El,c_\ell)\Bigr) \nonumber \\
&&\ \ \ \ \ \ \ \ \ \ \ \ \ \ \ \ \ \ \ \ \ \ \ \ \ \ \ \ \ \ \ \
+(1+10d_V)
\Bigl(\gcal_{-1}(\El,c_\ell)-\frac3{1+2r}\fcal_{-1}(\El,c_\ell)\Bigr)
\,\Bigr] \nonumber \\
&&\ \ \ \ \ \ \ \ \ \ \ \ \ \ \ \ \ \ \ \
-64d_R(d_V^2+d_A^2)^2
\Bigl[\,
\frac1{z}\Bigl(\gcal_0(\El,c_\ell)-\frac3{1+2r}\fcal_0(\El,c_\ell)\Bigr) \nonumber \\
&&\ \ \ \ \ \ \ \ \ \ \ \ \ \ \ \ \ \ \ \ \ \ \ \ \ \ \ \ \ \ \ \
-\gcal_{-1}(\El,c_\ell)+\frac3{1+2r}\fcal_{-1}(\El,c_\ell) \nonumber \\
&&\ \ \ \ \ \ \ \ \ \ \ \ \ \ \ \ \ \ \ \ \ \ \ \ \ \ \ \ \ \ \ \
-\frac4{z^2}\Bigl(\gcal_1(\El,c_\ell)-\frac3{1+2r}\fcal_1(\El,c_\ell)\Bigr)
\,\Bigr]
\,\biggr].
\end{eqnarray}
Here $\fcal_m$ and $\gcal_m\,(m=-2,-1,0,+1)$ are ${\mit\Omega}_t$ integrations choosing
the $\mib{p}_\ell$ direction as the $z$ axis
\begin{eqnarray*}
&&\fcal_m(\El,c_\ell)\equiv \int^{c_{t+}}_{c_{t-}} \!\!dc_t \int^{2\pi}_0 \!\!d\phi_t \,\omega v^m
\\ %,\ \ \ \
&&\gcal_m(\El,c_\ell)\equiv \int^{c_{t+}}_{c_{t-}} \!\!dc_t \int^{2\pi}_0 \!\!d\phi_t \,\frac{\omega}{1-\omega} v^m
\end{eqnarray*}
($c_t \equiv \cos\theta_t$) with
\begin{eqnarray}
&&c_{t+} = {\rm Max}\Bigl[\,{\rm Min}[\,\frac1{\hat{\beta}}
 \Bigl(1-\frac{M_W^2}{\sqrt{\hat{s}}E_\ell}\Bigr),\;+1\,],\;-1\,\Bigr]
\nonumber \\
&&c_{t-} = {\rm Min}\Bigl[\,{\rm Max}[\,\frac1{\hat{\beta}}
 \Bigl(1-\frac{m_t^2}{\sqrt{\hat{s}}E_\ell}\Bigr),\;-1\,],\;+1\,\Bigr]
\label{UpDown}
\end{eqnarray}
(see also \cite{Arens:1994jp}), and they are given as
\begin{equation}
\fcal_m = I_m(c_{t+})-I_m(c_{t-}),\ \ \ \ \ \gcal_m = J_m(c_{t+})-J_m(c_{t-}),
\end{equation}
where each $I_m$ and $J_m$ are
\begin{eqnarray}
&&I_1(c_t)=-\frac{\pi}{4(1-\hat{\beta})}
c_t \Bigl[\,(1-\hat{\beta}-x_\ell)
\Bigl(\hat{\beta}^2 s_\ell^2-2+\frac13\hat{\beta}^2(3c_\ell^2-1)c_t^2\Bigr)
\nonumber\\
&&\ \ \ \ \ \ \ \ \ \ \ \ \ \ \ \ \ \ \
+\frac14\hat{\beta} x_\ell [\, 2(\hat{\beta}^2 s_\ell^2-2)c_t +\hat{\beta}^2(3c_\ell^2-1)c_t^3\,]\,\Bigr] \\
&&I_0(c_t)=\frac{\pi}{1-\hat{\beta}} c_t[\,2(1-\hat{\beta}-x_\ell)+\hat{\beta} x_\ell c_t\,] \\
&&I_{-1}(c_t)
=\frac{4\pi}{1-\hat{\beta}}\Bigl[\, (1-\hat{\beta}-x_\ell)[\, f^+_{0/1}(c_t)+f^-_{0/1}(c_t) \,]
\nonumber\\
&&\ \ \ \ \ \ \ \ \ \ \ \ \ \ \ \ \ \ \
+\hat{\beta} x_\ell [\, f^+_{1/1}(c_t)+f^-_{1/1}(c_t) \,] \,\Bigr] \\
&&I_{-2}(c_t)
=\frac{8\pi}{1-\hat{\beta}}\Bigl[\, (1-\hat{\beta}-x_\ell)
[\, \hat{\beta} c_\ell f^+_{1/3}(c_t)+f^+_{0/3}(c_t)-\hat{\beta} c_\ell f^-_{1/3}(c_t)
\nonumber\\
&&\ \ \ \ \ \ \ \ \ \ \ \ \ \ \ \ \ \ \ \ \ \ \ \ \ \ \ \
+f^-_{0/3}(c_t) +f^+_{0/1}(c_t)+f^-_{0/1}(c_t) \,] \nonumber \\
&&\ \ \ \ \ \ \ \ \ \ \ \ \ \ \ \ \ \ \
+\hat{\beta} x_\ell
[\, \hat{\beta} c_\ell f^+_{2/3}(c_t)+f^+_{1/3}(c_t)-\hat{\beta} c_\ell f^-_{2/3}(c_t)+f^-_{1/3}(c_t)
\nonumber\\
&&\ \ \ \ \ \ \ \ \ \ \ \ \ \ \ \ \ \ \ \ \ \ \ \ \ \ \ \
+f^+_{1/1}(c_t)+f^-_{1/1}(c_t) \,] \,\Bigr] \\
&&J_1(c_t)
=\frac{\pi}{4}
 \Bigl[\,\frac1{\hat{\beta} x_\ell}(1-\hat{\beta})(\hat{\beta}^2-3) s_\ell^2 \ln(1-\hat{\beta} c_t)
%%%%% \nonumber\\
%%%%% &&\ \ \ \ \ \ \ \ \ \ \ \ \ \ \ \ \ \ \
+\frac1{2x_\ell}(1-\hat{\beta})(3c_\ell^2-1) c_t ( 2+\hat{\beta} c_t )
\nonumber \\
&&\ \ \ \ \ \ \ \ \ \ \ \ \ \ \ \ \ \ \
+(\hat{\beta}^2 s_\ell^2-2)c_t +\frac13\hat{\beta}^2(3c_\ell^2-1)c_t^3 \,\Bigr]
\\
&&J_0(c_t)
=-\frac{2\pi}{\hat{\beta}}
 \Bigl[\, \frac{1-\hat{\beta}}{x_\ell}\ln(1-\hat{\beta} c_t)+\hat{\beta} c_t \,\Bigr]
\\
&&J_{-1}(c_t)
= 4\pi \Bigl[\, \frac{1-\hat{\beta}}{x_\ell}[\, g_{0/1}^+(c_t)+g_{0/1}^-(c_t) \,]
-f_{0/1}^+(c_t)-f_{0/1}^-(c_t) \,\Bigr]
\\
&&J_{-2}(c_t)
= 8\pi \Bigl[\, \frac{1-\hat{\beta}}{x_\ell}
[\, \hat{\beta} c_\ell g_{1/3}^+(c_t)+g_{0/3}^+(c_t)-\hat{\beta} c_\ell g_{1/3}^-(c_t)+g_{0/3}^-(c_t)
\nonumber\\
&&\ \ \ \ \ \ \ \ \ \ \ \ \ \ \ \ \ \ \ \ \ \ \ \ \ \ \ \
+g_{0/1}^+(c_t)+g_{0/1}^-(c_t) \,]
\nonumber \\
&&\ \ \ \ \ \ \ \ \ \ \ \ \ \ \ \ \ \ \
-\hat{\beta} c_\ell f_{1/3}^+(c_t)-f_{0/3}^+(c_t)+\hat{\beta} c_\ell f_{1/3}^-(c_t)-f_{0/3}^-(c_t)
\nonumber\\
&&\ \ \ \ \ \ \ \ \ \ \ \ \ \ \ \ \ \ \ \ \ \ \ \ \ \ \ \
-f_{0/1}^+(c_t)-f_{0/1}^-(c_t) \,\Bigr]
\end{eqnarray}
with $x_\ell\equiv
 2E_\ell\sqrt{(1-\hat{\beta})/(1+\hat{\beta})}/m_t$, and $f_{m/n}^{\pm}$ and $g_{m/n}^{\pm}$ being
\begin{eqnarray}
&&f_{m/n}^{\pm}(c_t) \equiv 
\int dc_t \frac{c_t^m}{\sqrt{(\hat{\beta}^2 c_t^2 \pm 2\hat{\beta} c_\ell c_t +1-\hat{\beta}^2 s_\ell^2)^n}}
\label{fmn}
\\
&&g_{m/n}^{\pm}(c_t) \equiv 
\int dc_t
 \frac{c_t^m}{(1-\hat{\beta} c_t)\sqrt{(\hat{\beta}^2 c_t^2 \pm 2\hat{\beta} c_\ell c_t +1-\hat{\beta}^2 s_\ell^2)^n}}\,.
\label{gmn}
\end{eqnarray}
Their explicit forms after the integrations are
\begin{eqnarray}
&&f_{0/1}^{\pm}(c_t)=\frac1{\hat{\beta}}\ln[\,\hat{\beta} c_t \pm c_\ell + R_{\pm}(c_t)\,] \\
&&f_{1/1}^{\pm}(c_t)
=\frac1{\hat{\beta}^2} \Bigl[\, R_{\pm} \mp c_\ell \ln[\,\hat{\beta} c_t \pm c_\ell + R_{\pm}(c_t)\,] \,\Bigr] \\
&&f_{0/3}^{\pm}(c_t)=\frac{\hat{\beta} c_t \pm c_\ell}{\hat{\beta}(1-\hat{\beta}^2)s_\ell^2 R_{\pm}(c_t)} \\
&&f_{1/3}^{\pm}(c_t)
=-\frac{1-\hat{\beta}^2 s_\ell^2 \pm \hat{\beta} c_\ell c_t}{\hat{\beta}^2(1-\hat{\beta}^2)s_\ell^2 R_{\pm}(c_t)} \\
&&f_{2/3}^{\pm}(c_t)=\frac1{\hat{\beta}^3}
\Bigl[\, \frac{\hat{\beta}(2c_\ell^2-1+\hat{\beta}^2 s_\ell^2)c_t\pm c_\ell(1-\hat{\beta}^2 s_\ell^2)}
{(1-\hat{\beta}^2)s_\ell^2 R_{\pm}(c_t)}
\nonumber \\
&&\ \ \ \ \ \ \ \ \ \ \ \ \ \ \ \ \ \ \
+ \ln[\,\hat{\beta} c_t \pm c_\ell + R_{\pm}(c_t)\,] \,\Bigr] \\
&&g_{0/1}^{\pm}(c_t)
=-\frac1{\hat{\beta} Q_{\pm}}
\ln\Big[\, \frac{1-\hat{\beta} c_t}{(1\pm c_\ell)(1-\hat{\beta}^2\pm\hat{\beta}^2 c_\ell +\hat{\beta} c_t)
 +Q_{\pm} R_{\pm}(c_t)} \,\Bigr] \\
&&g_{0/3}^{\pm}(c_t)
=-\frac{1\mp 2c_\ell-\hat{\beta}^2(1 \mp c_\ell)
 -\hat{\beta} c_t}{\hat{\beta} s_\ell^2[\,2-\hat{\beta}^2(3\mp c_\ell)+\hat{\beta}^4(1\mp c_\ell)\,]R_{\pm}(c_t)}
\nonumber \\
% &&\ \ \ \ \ \ \ \ \ \ \ \ \ \ \ \ \ \ \
&&\ \ \ \ \ \ \ \ \ \ \
-\frac1{\hat{\beta} Q_{\pm}^3}
\ln\Big[\, \frac{1-\hat{\beta} c_t}{(1\pm c_\ell)(1-\hat{\beta}^2\pm\hat{\beta}^2 c_\ell +\hat{\beta} c_t)
 +Q_{\pm} R_{\pm}(c_t)} \,\Bigr] \\
&&g_{1/3}^{\pm}(c_t)
=\frac1{\hat{\beta}} [\, g_{0/3}^{\pm}(c_t) - f_{0/3}^{\pm}(c_t) \,],
\end{eqnarray}
where
\begin{eqnarray*}
&&R_{\pm}(c_t) \equiv \sqrt{\hat{\beta}^2 c_t^2 \pm 2\hat{\beta} c_\ell c_t +1-\hat{\beta}^2 s_\ell^2},\ \ \
Q_{\pm} \equiv \sqrt{2 \pm 2 c_\ell-\hat{\beta}^2 s_\ell^2}
\end{eqnarray*}
and all terms which cancel out in $f_{m/n}^{\pm}(c_{t+})-f_{m/n}^{\pm}(c_{t-})$ and
$g_{m/n}^{\pm}(c_{t+})-g_{m/n}^{\pm}(c_{t-})$ have been dropped from the beginning, although
the right-hand sides of eqs.(\ref{fmn},\ref{gmn}), the definition of $f_{m/n}^{\pm}$ and $g_{m/n}^{\pm}$,
mean indefinite integrals.

% 444444444444444444444444444444444444444444444444444444444444
\sec{Final-lepton distributions}

We are now in the final stage of computing the lepton-momentum distributions under actual
experimental conditions. 
In order to derive hadron cross sections based on the parton-level formulas given in the
previous section, we first need to connect partonic cross sections in the parton-CM frame
and hadron-CM frame. The final-lepton energy and scattering angle in the parton-CM frame,
$E_{\ell}^*$ and $\theta_{\ell}^*$, are expressed in terms of those in the hadron-CM frame,
$\El$ and $\theta_{\ell}$, as
\begin{eqnarray}
&&E_{\ell}^*=\El (1 - \beta_{\rm L} c_{\ell})/\sqrt{1-\beta_{\rm L}^2}, \ \ \
c_{\ell}^*=(c_{\ell} - \beta_{\rm L} )/(1 - \beta_{\rm L} c_{\ell}),
\label{costhetat}
\end{eqnarray}
where $\beta_{\rm L}$ is the Lorentz-transformation boost factor connecting the two frames. These
relations lead to Jacobian
\begin{equation}
\partial(E_{\ell}^*,\,c_{\ell}^*)/\partial(\El,\,c_{\ell})
= \El/E_{\ell}^*
\end{equation}
and consequently cross-section relation
\begin{equation}
\frac{d\sigma_{q\bar{q},gg}}{d\El dc_{\ell}}
=\frac{\El}{E_{\ell}^*}
\frac{d\sigma_{q\bar{q},gg}}{dE_{\ell}^* dc_{\ell}^*}\,.
\end{equation}

Then the hadron cross sections are obtained by integrating the product of
{\it the parton-distribution functions} and {\it the parton cross sections in
the hadron-CM frame} on the momentum
fractions $x_1$ and $x_2$ carried by the partons:
\begin{equation}
\frac{d\sigma_{p\bar{p}/pp}}{d\El dc_{\ell}}
= \sum_{a,b} \int^1_{4m_t^2/s}\! \!\!\!dx_1 \int^1_{4m_t^2/(x_1 s)} \!\!\!\!\!\!\!\!dx_2
\:N_a(x_1) N_b(x_2)
\frac{\El}{E_{\ell}^*}
\frac{d\sigma_{ab}}{dE_{\ell}^* dc_{\ell}^*},
\end{equation}
where $N_{a,b}(x)$ are the parton-distribution functions of parton $a$ and $b$
($a,b=u,\bar{u}$, $d,\bar{d}$, $s,\bar{s}$, $c,\bar{c}$, $b,\bar{b}$ or $g$)
and the boost factor is given by $\beta_{\rm L} = (x_1-x_2)/(x_1+x_2)$. Note here that
$s$ is defined via the initial hadron momenta $p_{p/\bar{p}}$ as $s \equiv (p_p + p_{p/\bar{p}})^2$.

\subsec{Angular distribution}

\begin{figure}[H]
\vspace*{-0.4cm}
\begin{minipage}{14.8cm}
\begin{center}
% \hspace*{1.5cm}
%%% \psfrag{dv}{\begin{large}\hspace*{-0.0cm}$d_V$\end{large}}
%%% \psfrag{da}{\begin{large}\hspace*{-0.0cm}$d_A$\end{large}}
%%% \includegraphics[width=12cm]{allowed_index.eps}
\begin{overpic}[width=13cm,clip]{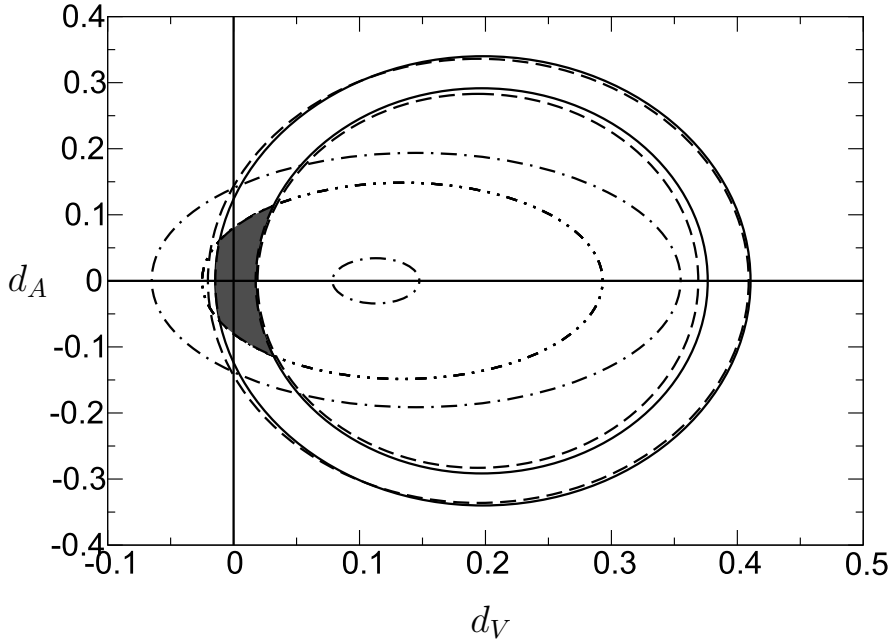}
\put(5,139){\large $d_A$}
% \put(8,120){\large $d_A$}
\put(180,10){\large $d_V$}
% \put(150,10){\large $d_V$}
\end{overpic}
\vspace*{-0.5cm}
\caption{The $d_{V,A}$ region allowed by Tevatron and LHC data altogether (the shaded part).
The solid curves, the dashed curves and the dash-dotted curves are respectively from CDF, D0
and CMS data, and the dash-dot-dotted curve is from ATLAS data.}\label{allowed}
\end{center}
\end{minipage}
\end{figure}

We first study the angular distribution:
\begin{equation}
\frac{d\sigma_{p\bar{p}/pp}}{dc_{\ell}}
= \int^{E_{\ell}^+}_{E_{\ell}^-} \!d\El \frac{d\sigma_{p\bar{p}/pp}}{d\El dc_{\ell}},
\label{Ang-Int}
\end{equation}
where
\[
E_{\ell}^+ = \frac{m_t^2}{\sqrt{s}(1-\beta)},\ \ \
E_{\ell}^- = \frac{M_W^2}{\sqrt{s}(1+\beta)}
\]
and $\beta \equiv \sqrt{1-4m_t^2/s}$. Concerning the anomalous-coupling parameters, we take
\[
(d_V,\:d_A)=\:({\rm a})\ (-0.01,\:0),\ \ \
            \:({\rm b})\ (0.01,\:0),\ \ \
            \:({\rm c})\ (0,\:0.05),\ \ \
            \:({\rm d})\ (0.03,\:0.10)
\]
as typical examples,\footnote{As you find in the appendix, these values are not excluded
    by the current experimental data (see Fig.\ref{allowed}). From now on, we only use
    such values as typical parameter sets.}\
and we use $\alpha_s =0.118$, $M_W=80.4$ GeV, $B_{\ell}=0.22\,(\ell=e/\mu)$, and the present
world average $m_t=172$ GeV \cite{:2008vn}.
Note that the decay anomalous parameter $d_R$ does not contribute to the angular
distribution due to the decoupling theorem \cite{Grzadkowski:1999iq}--\cite{Godbole:2006tq}.
As for the parton-distribution functions, we adopt the latest set ``CTEQ6.6M"
in Next-to-Next-to-Leading-Order (NNLO) approximation \cite{Nadolsky:2008zw}.

\begin{figure}[H]
\begin{minipage}{14.8cm}
\begin{center}
\psfrag{dsigcl}{\begin{Large}\hspace*{-0.80cm}
$\frac1{\sigma_{\scriptscriptstyle\rm SM}}\frac{d\sigma}{dc_{\ell}}$\end{Large}}
\psfrag{sm}{\begin{small}Standard model\end{small}}
\psfrag{dvm}{\begin{small}(a)$\:d_V=-0.01,\:d_A=0$\end{small}}
\psfrag{dv}{\begin{small}(b)$\:d_V=0.01,\:d_A=0$\end{small}}
\psfrag{da}{\begin{small}(c)$\:d_V=0,\:d_A=0.05$\end{small}}
\psfrag{dva}{\begin{small}(d)$\:d_V=0.03,\:d_A=0.10$\end{small}}
\psfrag{cl}{\begin{large}\hspace*{-0.2cm}$c_{\ell}(\equiv\cos \theta_\ell)$\end{large}}
\vspace*{0.5cm}
\hspace*{-3cm}
\includegraphics[width=12cm, origin=c, angle=-90]{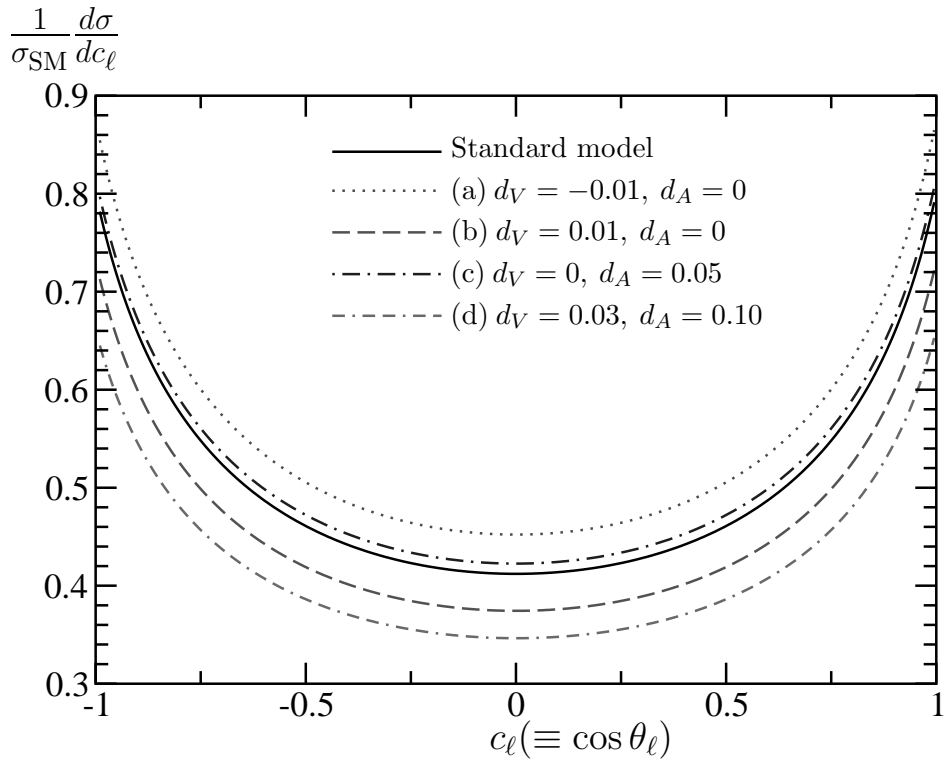}
\vspace*{-2.4cm}
\caption{The final-lepton angular distribution normalized
by $\sigma_{\rm SM}$: Tevatron energy $\sqrt{s}=1.96$ TeV}\label{tevcl}
\end{center}
\end{minipage}
\end{figure}

\begin{figure}[H]
\vspace*{-0.2cm}
\begin{minipage}{14.8cm}
\begin{center}
\psfrag{dsigcl}{\begin{Large}\hspace*{-0.8cm}
$\frac1{\sigma_{\small\rm SM}}\frac{d\sigma}{dc_{\ell}}$\end{Large}}
\psfrag{sm}{\begin{small}Standard model\end{small}}
\psfrag{dvm}{\begin{small}(a)$\:d_V=-0.01,\:d_A=0$\end{small}}
\psfrag{dv}{\begin{small}(b)$\:d_V=0.01,\:d_A=0$\end{small}}
\psfrag{da}{\begin{small}(c)$\:d_V=0,\:d_A=0.05$\end{small}}
\psfrag{dva}{\begin{small}(d)$\:d_V=0.03,\:d_A=0.10$\end{small}}
\psfrag{cl}{\begin{large}\hspace*{-0.2cm}$c_{\ell}(\equiv\cos \theta_\ell)$\end{large}}
\vspace*{0.5cm}
\hspace*{-3cm}
\includegraphics[width=12cm, origin=c, angle=-90]{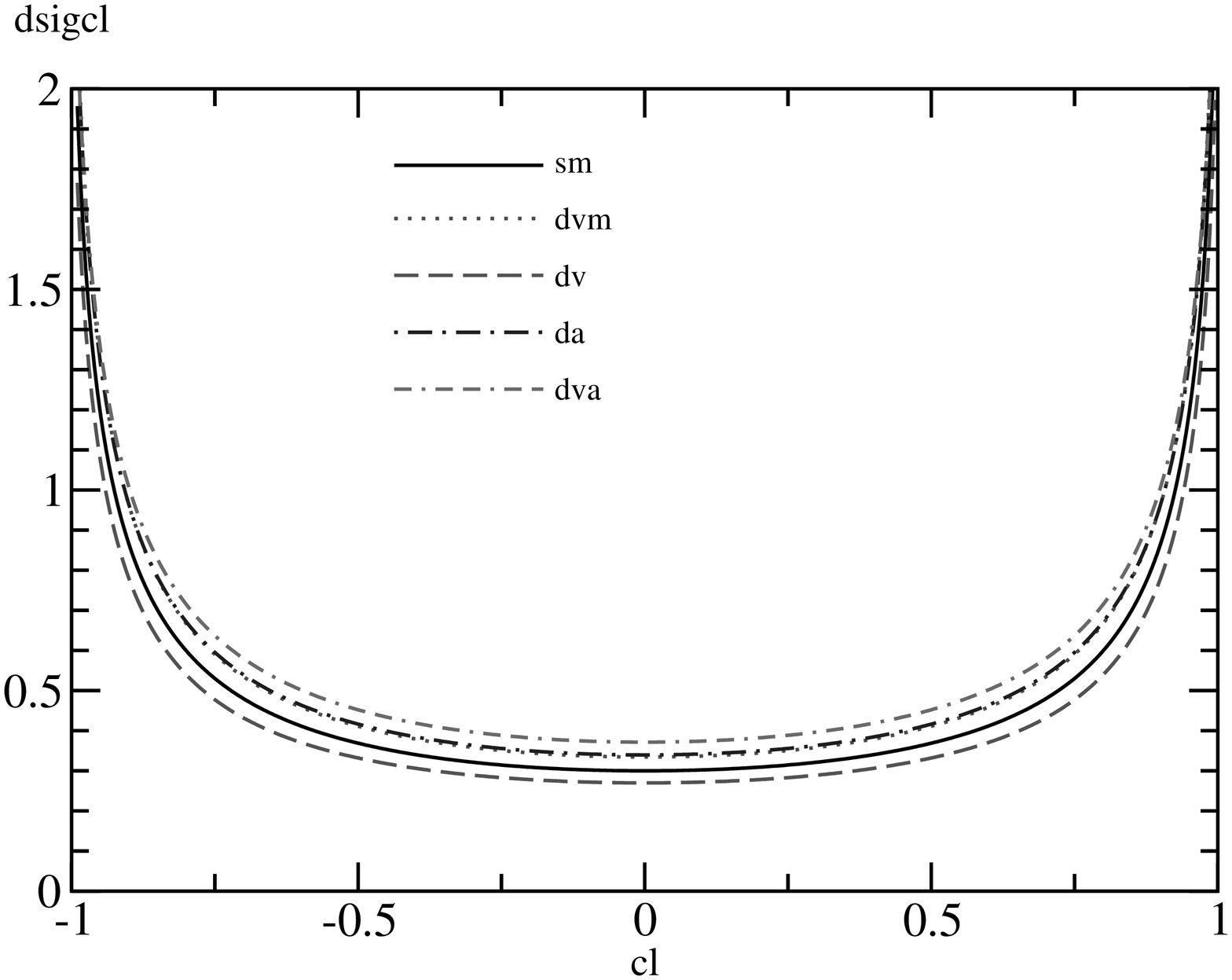}
\vspace*{-2.4cm}
\caption{The final-lepton angular distribution normalized
by $\sigma_{\rm SM}$: LHC energy $\sqrt{s}=7$ TeV}\label{lhccl7}
\end{center}
\end{minipage}
\end{figure}

\begin{figure}[H]
\vfill % \vspace*{0.4cm}
\begin{minipage}{14.8cm}
\begin{center}
\psfrag{dsigcl}{\begin{Large}\hspace*{-0.8cm}
$\frac1{\sigma_{\rm SM}}\frac{d\sigma}{dc_{\ell}}$\end{Large}}
\psfrag{sm}{\begin{small}Standard model\end{small}}
\psfrag{dvm}{\begin{small}(a)$\:d_V=-0.01,\:d_A=0$\end{small}}
\psfrag{dv}{\begin{small}(b)$\:d_V=0.01,\:d_A=0$\end{small}}
\psfrag{da}{\begin{small}(c)$\:d_V=0,\:d_A=0.05$\end{small}}
\psfrag{dva}{\begin{small}(d)$\:d_V=0.03,\:d_A=0.10$\end{small}}
\psfrag{cl}{\begin{large}\hspace*{-0.2cm}$c_{\ell}(\equiv\cos \theta_\ell)$\end{large}}\hspace*{1cm}
% \vspace*{0.9cm} 
\hspace*{-4.2cm} % \hspace*{-3cm}
\includegraphics[width=12cm, origin=c, angle=-90]{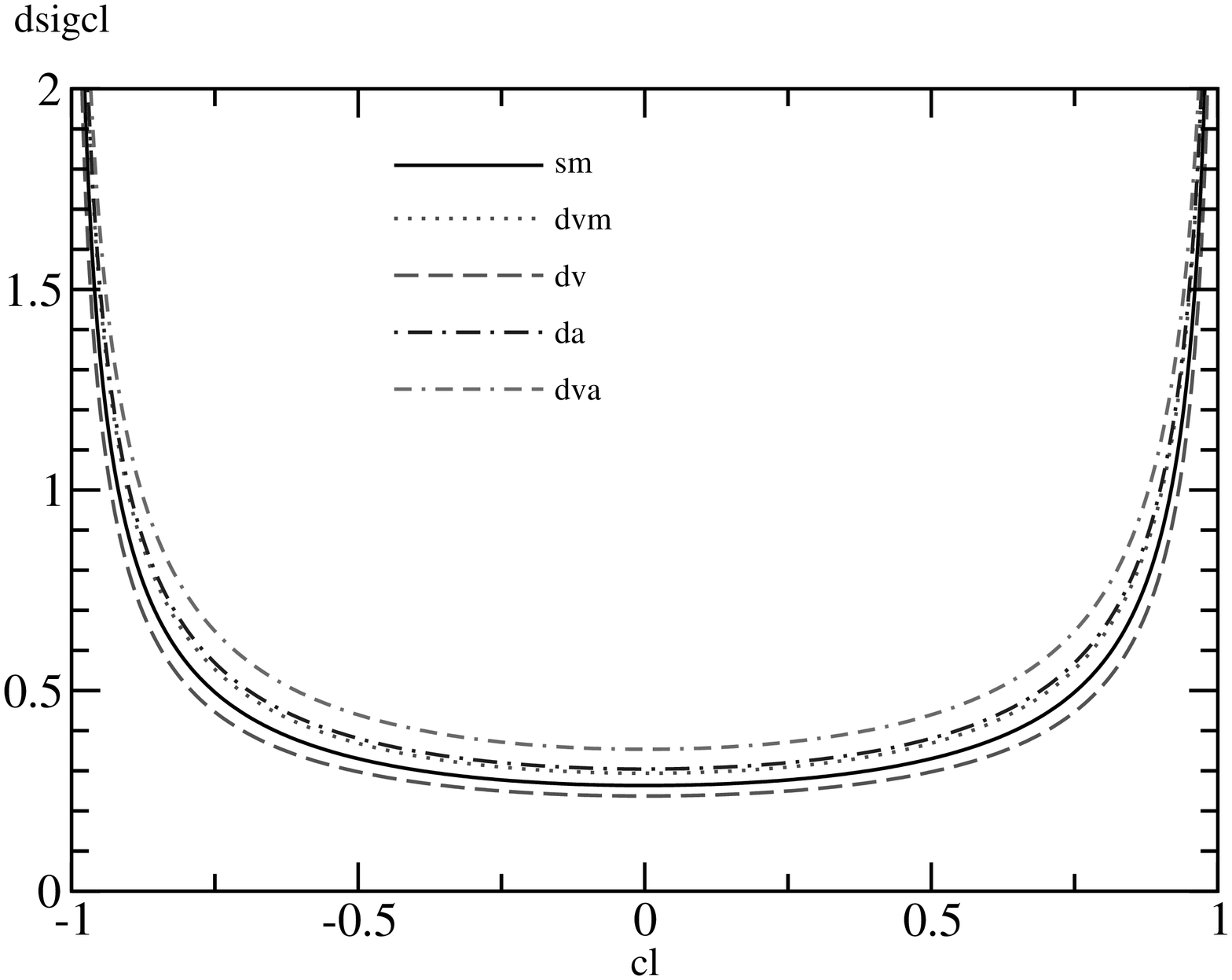}
\vspace*{-2.4cm}
\caption{The final-lepton angular distribution normalized
by $\sigma_{\rm SM}$: LHC energy $\sqrt{s}=14$ TeV}\label{lhccl14}
\end{center}
\end{minipage}
\end{figure}

We present the results in Fig.\ref{tevcl}--Fig.\ref{lhccl14} for Tevatron, LHC(7 TeV) and LHC(14 TeV)
respectively, where we show the distributions normalized by the standard-model total cross section
$\sigma_{\rm SM} \equiv \sigma(p\bar{p}/pp \to t\bar{t}X \to \ell^+ X')
=B_{\ell} \sigma(p\bar{p}/pp \to t\bar{t}X)$ so that large part of the QCD corrections
cancel each other in the ratio.\footnote{Strictly speaking, of course, the QCD corrections
    to the total cross sections and differential cross sections are not the same as each
    other, but the difference is not that sizable as studied systematically 
    in \cite{Bevilacqua:2010qb} (see also \cite{Frederix:2007gi}).}\ 
It should be noted that the vertical axis of Fig.\ref{tevcl} is different from those of
Fig.\ref{lhccl7} and Fig.\ref{lhccl14} in scale.

Through those figures, we find that the deviation from $d\sigma_{\rm SM}$ varies to a certain extent
to positive or negative direction depending on the anomalous parameters, even if we strictly take into account
the constraints on $d_{V,A}$ coming from combined Tevatron and LHC data shown in the appendix and
change parameter values only within the resultant allowed region. In some cases, it will not be easy
to distinguish the curves: In particular, those with parameter sets (a) and (c) almost overlap
each other in Figs.\ref{lhccl7} and \ref{lhccl14}.

\vfill % \vskip 1cm

% \newpage

%%%%%%%%%%%%%%%%%%%%%%%%%%%%%%%%%%%%%%%%%%%%%%%%%%%%%%%%%%%%%%%%%%
%
%   tevcl.eps
%
%%%%%%%%%%%%%%%%%%%%%%%%%%%%%%%%%%%%%%%%%%%%%%%%%%%%%%%%%%%%%%%%%%%%%

\newpage
%%%%%%%%%%%%%%%%%%%%%%%%%%%%%%%%%%%%%%%%%%%%%%%%%%%%%%%%%%%%%%%%%%%%%%%%
%
%
%%%%%%%%%%%%%%%%%%%%%%%%%%%%%%%%%%%%%%%%%%%%%%%%%%%%%%%%%%%%%%%%%%
%
%   lhccl7.eps
%
%%%%%%%%%%%%%%%%%%%%%%%%%%%%%%%%%%%%%%%%%%%%%%%%%%%%%%%%%%%%%%%%%%%%%
% \newpage 
%%%%%%%%%%%%%%%%%%%%%%%%%%%%%%%%%%%%%%%%%%%%%%%%%%%%%%%%%%%%%%%%%%%%%%%%
%
%
%%%%%%%%%%%%%%%%%%%%%%%%%%%%%%%%%%%%%%%%%%%%%%%%%%%%%%%%%%%%%%%%%%
%
%   lhccl14.eps
%
%%%%%%%%%%%%%%%%%%%%%%%%%%%%%%%%%%%%%%%%%%%%%%%%%%%%%%%%%%%%%%%%%%%%%
%%%%%%%%%%%%%%%%%%%%%%%%%%%%%%%%%%%%%%%%%%%%%%%%%%%%%%%%%%%%%%%%%%%%%

% \vskip 0.8cm

In order to clarify the size of those nonstandard effects more quantitatively, let us show
\[
\delta(*)\equiv [\,d\sigma(*)-d\sigma_{\rm SM}\,]/d\sigma_{\rm SM}\,(\times 100)
\]
at $\cos\theta_{\ell}=0$ as an example, where $*$ means parameter set (a), (b), (c) or (d):
\underline{Tevatron}
\begin{equation}
\begin{array}{ll}
\delta({\rm a}) = +\ 9.73\,\%,~~~~~~& \delta({\rm b}) = -\ 9.15\,\%, \\
\delta({\rm c}) = +\ 2.54\,\%,      & \delta({\rm d}) = -15.94\,\%. \\
\end{array}
\end{equation}
\underline{LHC} (7 TeV)
\begin{equation}
\begin{array}{ll}
\delta({\rm a}) = +11.50\,\%,~~~~~~& \delta({\rm b}) = -10.02\,\%, \\
\delta({\rm c}) = +13.20\,\%,      & \delta({\rm d}) = +23.82\,\%. \\
\end{array}
\end{equation}
\underline{LHC} (14 TeV)
\begin{equation}
\begin{array}{ll}
\delta({\rm a}) = +11.61\,\%,~~~~~~& \delta({\rm b}) = -\ 9.95\,\%, \\
\delta({\rm c}) = +15.60\,\%,      & \delta({\rm d}) = +34.31\,\%. \\
\end{array}
\end{equation}
The deviation could be as large as more than 30 \% at LHC, and there seem to be some
chances of getting a nonstandard signal.

\subsec{Energy distribution}

\begin{figure}[H]
\vspace*{-0.2cm}
\begin{minipage}{14.8cm}
\begin{center}
\psfrag{dsigel}{\begin{Large}\hspace*{-0.50cm}
$\frac1{\sigma_{\rm SM}}\frac{d\sigma}{d E_\ell}$\end{Large}[GeV$^{-1}$]}
\psfrag{sm}{\begin{small}Standard model\end{small}}
\psfrag{dvm}{\begin{small}(a)$\:d_V=-0.01,\:d_A=0,\:d_R=0$\end{small}}
\psfrag{dv}{\begin{small}(b)$\:d_V=0.01,\:d_A=0,\:d_R=0$\end{small}}
\psfrag{da}{\begin{small}(c)$\:d_V=0,\:d_A=0.05,\:d_R=0$\end{small}}
\psfrag{dr}{\begin{small}(d)$\:d_V=0,\:d_A=0,\:d_R=0.01$\end{small}}
\psfrag{dvar}{\begin{small}(e)$\:d_V=0.03,\:d_A=0.10,\:d_R=0.01$\end{small}}
\psfrag{el}{\begin{large}\hspace*{-0.2cm}$E_\ell $\end{large}}
\psfrag{gev}{\hspace*{-0.2cm}[GeV]}
%% \vspace*{-0.5cm}
\hspace*{-3cm}
\includegraphics[width=13cm, origin=c, angle=-90]{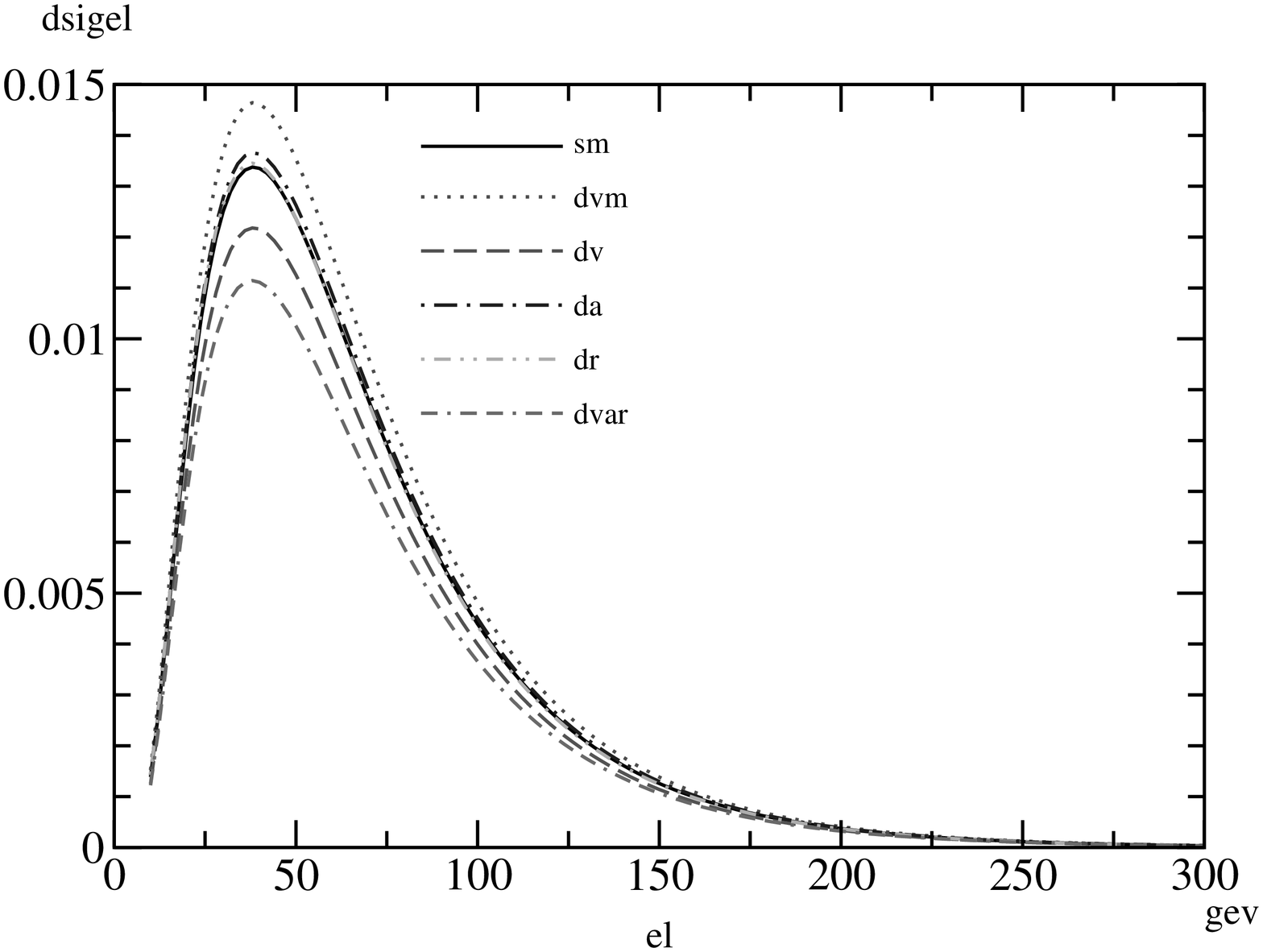}
\vspace*{-2.8cm}
\caption{The final-lepton energy distribution normalized
by $\sigma_{\rm SM}$: Tevatron energy $\sqrt{s}=1.96$ TeV}\label{tevel}
%\end{center}
%\end{minipage}
%\end{figure}
%%%%%%%%%%%%%%%%%%%%%%%%%%%%%%%%%%%%%%%%%%%%%%%%%%%%%%%%%%%%%%%%%%%%%%%%
%
%
%%%%%%%%%%%%%%%%%%%%%%%%%%%%%%%%%%%%%%%%%%%%%%%%%%%%%%%%%%%%%%%%%%
%
%   lhcel7.eps
%
%%%%%%%%%%%%%%%%%%%%%%%%%%%%%%%%%%%%%%%%%%%%%%%%%%%%%%%%%%%%%%%%%%%%%
%\begin{figure}[H]
%\begin{minipage}{14.8cm}
%\begin{center}
%\vspace*{0.5cm}
\psfrag{dsigel}{\begin{Large}\hspace*{-0.80cm}
$\frac1{\sigma_{\rm SM}}\frac{d\sigma}{d E_\ell}$\end{Large}[GeV$^{-1}$]}
\psfrag{sm}{\begin{small}Standard model\end{small}}
\psfrag{dvm}{\begin{small}(a)$\:d_V=-0.01,\:d_A=0,\:d_R=0$\end{small}}
\psfrag{dv}{\begin{small}(b)$\:d_V=0.01,\:d_A=0,\:d_R=0$\end{small}}
\psfrag{da}{\begin{small}(c)$\:d_V=0,\:d_A=0.05,\:d_R=0$\end{small}}
\psfrag{dr}{\begin{small}(d)$\:d_V=0,\:d_A=0,\:d_R=0.01$\end{small}}
\psfrag{dvar}{\begin{small}(e)$\:d_V=0.03,\:d_A=0.10,\:d_R=0.01$\end{small}}
\psfrag{el}{\begin{large}\hspace*{-0.2cm}$E_\ell $\end{large}}
\psfrag{gev}{\hspace*{-0.2cm}[GeV]}
\vspace*{0.5cm}
\hspace*{-3cm}
\includegraphics[width=13cm, origin=c, angle=-90]{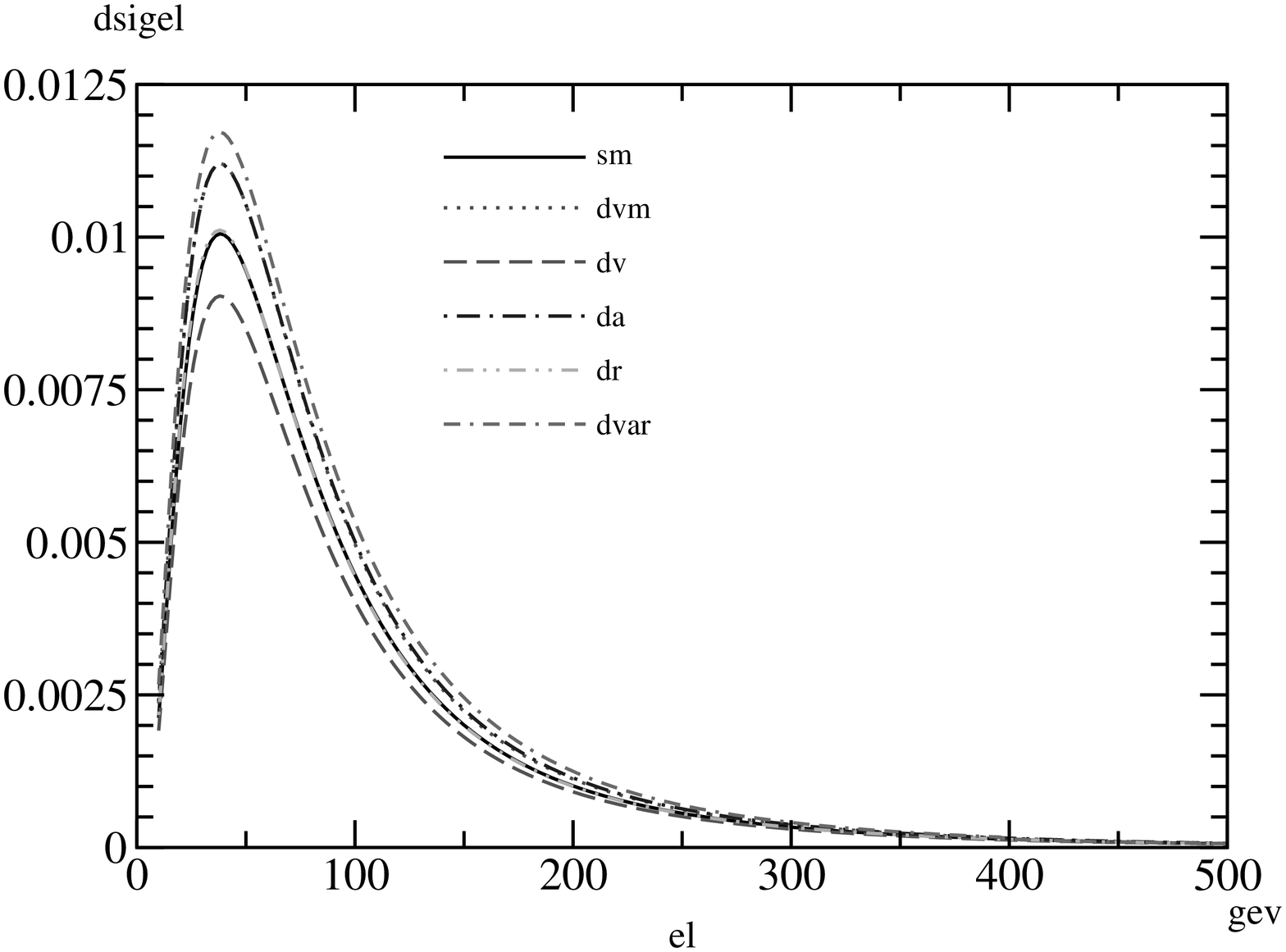}
\vspace*{-2.8cm}
\caption{The final-lepton energy distribution normalized
by $\sigma_{\rm SM}$: LHC energy $\sqrt{s}=7$ TeV}\label{lhcel7}
\end{center}
\end{minipage}
\end{figure}

\begin{figure}[H]
\vspace*{-0.4cm}
\begin{minipage}{14.8cm}
\begin{center}
\psfrag{dsigel}{\begin{Large}\hspace*{-0.80cm}
$\frac1{\sigma_{\rm SM}}\frac{d\sigma}{d E_\ell}$\end{Large}[GeV$^{-1}$]}
\psfrag{sm}{\begin{small}Standard model\end{small}}
\psfrag{dvm}{\begin{small}(a)$\:d_V=-0.01,\:d_A=0,\:d_R=0$\end{small}}
\psfrag{dv}{\begin{small}(b)$\:d_V=0.01,\:d_A=0,\:d_R=0$\end{small}}
\psfrag{da}{\begin{small}(c)$\:d_V=0,\:d_A=0.05,\:d_R=0$\end{small}}
\psfrag{dr}{\begin{small}(d)$\:d_V=0,\:d_A=0,\:d_R=0.01$\end{small}}
\psfrag{dvar}{\begin{small}(e)$\:d_V=0.03,\:d_A=0.10,\:d_R=0.01$\end{small}}
\psfrag{el}{\begin{large}\hspace*{-0.2cm}$E_\ell $\end{large}}
\psfrag{gev}{\hspace*{-0.2cm}[GeV]}
\vspace*{0.5cm}
\hspace*{-3cm}
\includegraphics[width=13cm, origin=c, angle=-90]{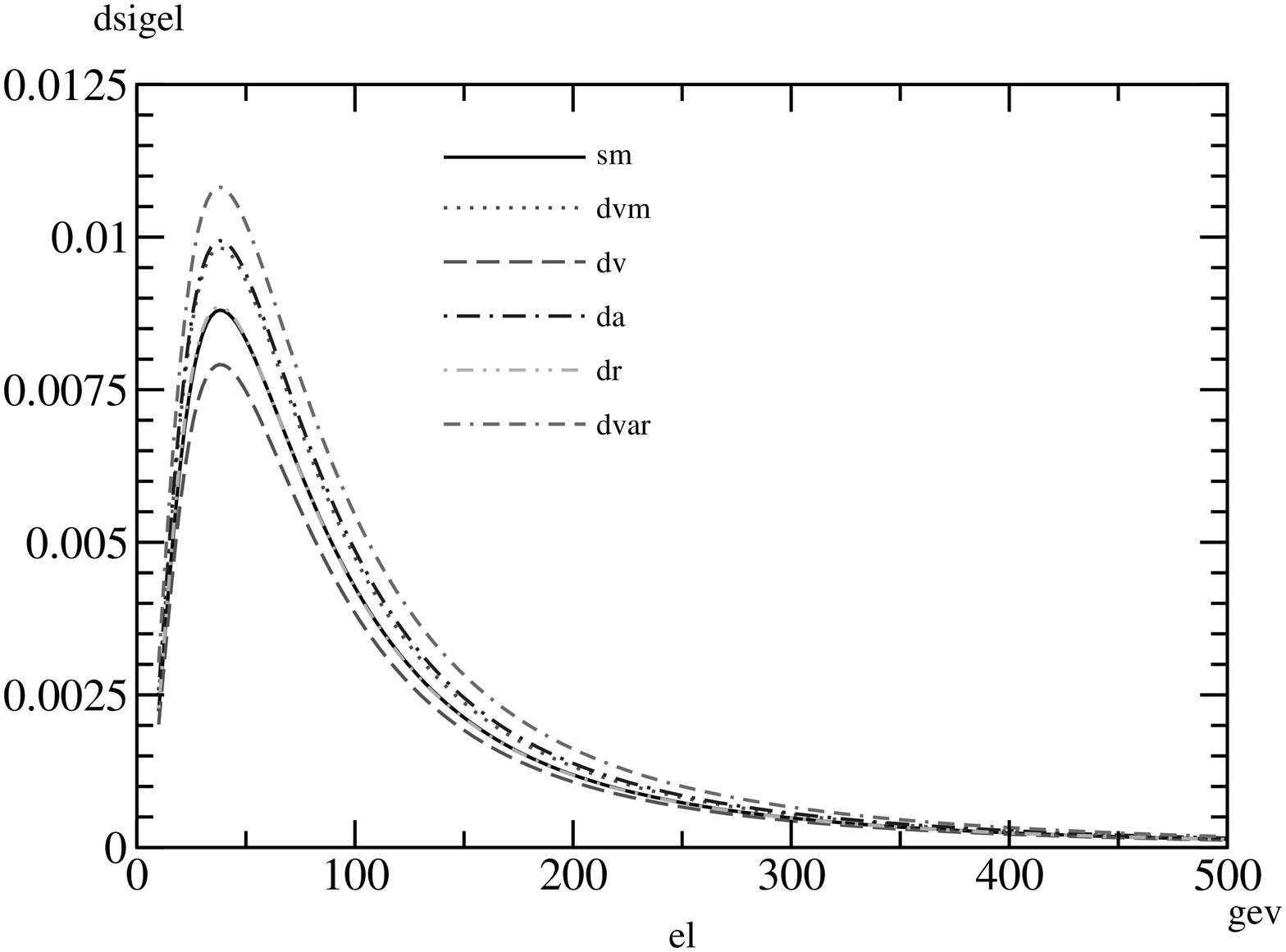}
\vspace*{-2.8cm}
\caption{The final-lepton energy distribution normalized
by $\sigma_{\rm SM}$: LHC energy $\sqrt{s}=14$ TeV}\label{lhcel14}
\end{center}
\end{minipage}
\end{figure}

Let us next study the energy distribution:
\begin{equation}
\frac{d\sigma_{p\bar{p}/pp}}{d\El}
= \int^{c_{\ell}^+}_{c_{\ell}^-} \!dc_{\ell} \frac{d\sigma_{p\bar{p}/pp}}{d\El dc_{\ell}},
\end{equation}
where
\[
c_{\ell}^+ = +1\,,\ \ \
c_{\ell}^- = -1\,.
\]
In the same way as the angular distributions,
we show the normalized distributions in Figs.\ref{tevel}--\ref{lhcel14}
using anomalous-coupling parameters:
\begin{eqnarray*}
&&(d_V,\:d_A,\:d_R)=\:({\rm a})\ (-0.01,\:0,\:0),\ \ \
                          ({\rm b})\ (0.01,\:0,\:0),\ \ \
                          ({\rm c})\ (0,\:0.05,\:0),\\
&&\phantom{(d_V,\:d_A,\:d_R)=\:}\ 
                          ({\rm d})\ (0,\:0,\:0.01),\phantom{-}\ \ \
                          ({\rm e})\ (0.03,\:0.10,\:0.01).
\end{eqnarray*}
Note that the $d_R$ terms can also contribute to the results in this case.

We see that sizable effects can be expected in some cases. However, Figs.\ref{tevel}--\ref{lhcel14}
tell us that the dash-dot-dotted curve
depicted with parameter set (d) has a substantial overlap with the SM curve, which
indicates that we have little chance to observe any signal in this case. In addition,
those with parameter sets (a) and (c) show quite similar behavior
at LHC and will be indistinguishable from each other, though the overlapping part gets
smaller as the center-of-mass energy increases.

% \vfill

%%%%%%%%%%%%%%%%%%%%%%%%%%%%%%%%%%%%%%%%%%%%%%%%%%%%%%%%%%%%%%%%%%
%
%   tevel.eps
%
%%%%%%%%%%%%%%%%%%%%%%%%%%%%%%%%%%%%%%%%%%%%%%%%%%%%%%%%%%%%%%%%%%%%%
%%%%%%%%%%%%%%%%%%%%%%%%%%%%%%%%%%%%%%%%%%%%%%%%%%%%%%%%%%%%%%%%%%%%%%%%
%
%
%%%%%%%%%%%%%%%%%%%%%%%%%%%%%%%%%%%%%%%%%%%%%%%%%%%%%%%%%%%%%%%%%%
%
%   lhcel14.eps
%
%%%%%%%%%%%%%%%%%%%%%%%%%%%%%%%%%%%%%%%%%%%%%%%%%%%%%%%%%%%%%%%%%%%%%
%%%%%%%%%%%%%%%%%%%%%%%%%%%%%%%%%%%%%%%%%%%%%%%%%%%%%%%%%%%%%%%%%%%%%

\vskip 0.8cm
\noindent
We again show the size of the deviations
in percentage at $E_\ell=50$ GeV:
\\
\underline{Tevatron}
\begin{equation}
\begin{array}{lll}
\delta({\rm a}) = +9.48\,\%,\phantom{0}~~~~~~& \delta({\rm b}) = -8.94\,\%,~~~~~~& \delta({\rm c}) = +2.13\,\%,\phantom{0} \\
\delta({\rm d}) = +0.02\,\%,      & \delta({\rm e}) = -17.08\,\%.      &                         \\
\end{array}
\end{equation}
\underline{LHC} (7 TeV)
\begin{equation}
\begin{array}{lll}
\delta({\rm a}) = +11.46\,\%,~~~~~~& \delta({\rm b}) = -10.12\,\%,~~~~~~& \delta({\rm c}) = +11.45\,\%, \\
\delta({\rm d}) = +0.20\,\%,      & \delta({\rm e}) = +16.42\,\%.      &                         \\
\end{array}
\end{equation}
\underline{LHC} (14 TeV)
\begin{equation}
\begin{array}{lll}
\delta({\rm a}) = +11.58\,\%,~~~~~~& \delta({\rm b}) = -10.11\,\%,~~~~~~& \delta({\rm c}) = +13.02\,\%, \\
\delta({\rm d}) = +0.25\,\%,      & \delta({\rm e}) = +22.92\,\%.      &                         \\
\end{array}
\end{equation}
The size of the deviation is similar to that of the angular distribution, and it could be
fairly large though depending on the parameters.

\subsec{Transverse-momentum distribution}

Finally, we compute the transverse-momentum $p_{\rm T}$ distribution. This distribution
is obtained by integrating $d\sigma_{p\bar{p}/pp}/dp_{\rm T} dc_{\ell}$ over $c_{\ell}$,
which cross section is connected with $d\sigma_{p\bar{p}/pp}/dE_{\ell} dc_{\ell}$
through Jacobian $1/\sqrt{1-c_{\ell}^2}\,$:
\begin{equation}
\frac{d\sigma_{p\bar{p}/pp}}{dp_{\rm T}}
= \int^{c_{\ell}^+}_{c_{\ell}^-} \!dc_{\ell} \frac1{\sqrt{1-c_{\ell}^2}}\,
\frac{d\sigma_{p\bar{p}/pp}}{d\El dc_{\ell}},
\end{equation}
where
\[
c_{\ell}^+ = -c_{\ell}^- = \sqrt{1-(p_{\rm T}/E_{\ell}^+)^2}
\]
and $E_{\ell}^+$ is given in eq.(\ref{Ang-Int}).
Using the same anomalous-coupling parameters as for the energy distributions
\begin{eqnarray*}
&&(d_V,\:d_A,\:d_R)=\:({\rm a})\ (-0.01,\:0,\:0),\ \ \
                          ({\rm b})\ (0.01,\:0,\:0),\ \ \
                          ({\rm c})\ (0,\:0.05,\:0),\\
&&\phantom{(d_V,\:d_A,\:d_R)=\:}\ 
                          ({\rm d})\ (0,\:0,\:0.01),\phantom{-}\ \ \
                          ({\rm e})\ (0.03,\:0.10,\:0.01),
\end{eqnarray*}

%%%%%%%%%%%%%%%%%%%%%%%%%%%%%%%%%%%%%%%%%%%%%%%%%%%%%%%%%%%%%%%%%%
%
%   tevpt.eps
%
%%%%%%%%%%%%%%%%%%%%%%%%%%%%%%%%%%%%%%%%%%%%%%%%%%%%%%%%%%%%%%%%%%%%%
\begin{figure}[H]
\vfill
\begin{minipage}{14.8cm}
\begin{center}
\psfrag{dsigpt}{\begin{Large}\hspace*{-0.50cm}
$\frac1{\sigma_{\rm SM}}\frac{d\sigma}{d p_{\rm T}}$\end{Large}[GeV$^{-1}$]}
\psfrag{sm}{\begin{small}Standard model\end{small}}
\psfrag{dvm}{\begin{small}(a)$\:d_V=-0.01,\:d_A=0,\:d_R=0$\end{small}}
\psfrag{dv}{\begin{small}(b)$\:d_V=0.01,\:d_A=0,\:d_R=0$\end{small}}
\psfrag{da}{\begin{small}(c)$\:d_V=0,\:d_A=0.05,\:d_R=0$\end{small}}
\psfrag{dr}{\begin{small}(d)$\:d_V=0,\:d_A=0,\:d_R=0.01$\end{small}}
\psfrag{dvar}{\begin{small}(e)$\:d_V=0.03,\:d_A=0.10,\:d_R=0.01$\end{small}}
\psfrag{pt}{\begin{large}\hspace*{-0.2cm}$p_{\rm T} $\end{large}}
\psfrag{gev}{\hspace*{-0.2cm}[GeV]}
\hspace*{-3cm}
\includegraphics[width=13cm, origin=c, angle=-90]{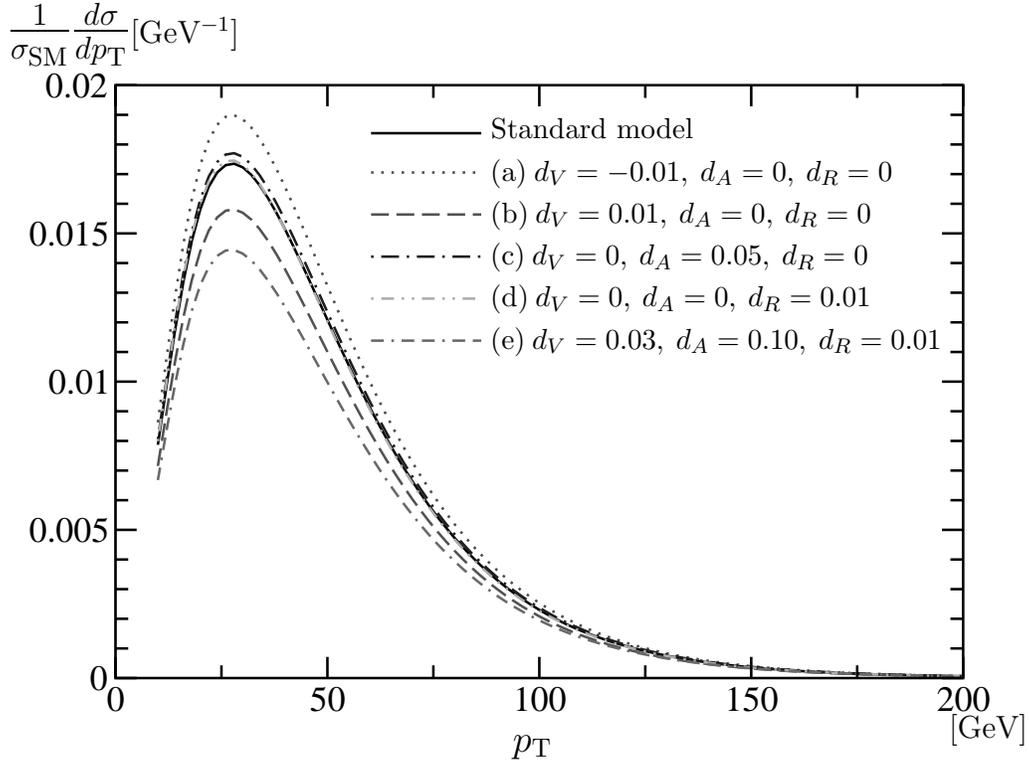}
\vspace*{-2.8cm}
\caption{The final-lepton transverse-momentum distribution normalized
by $\sigma_{\rm SM}$: Tevatron energy $\sqrt{s}=1.96$ TeV}\label{tevpt}
\end{center}
\end{minipage}
\end{figure}
%%%%%%%%%%%%%%%%%%%%%%%%%%%%%%%%%%%%%%%%%%%%%%%%%%%%%%%%%%%%%%%%%%%%%%%%
%
%
%%%%%%%%%%%%%%%%%%%%%%%%%%%%%%%%%%%%%%%%%%%%%%%%%%%%%%%%%%%%%%%%%%
%
%   lhcpt7.eps
%
%%%%%%%%%%%%%%%%%%%%%%%%%%%%%%%%%%%%%%%%%%%%%%%%%%%%%%%%%%%%%%%%%%%%%
\newpage
\begin{figure}[H]
\begin{minipage}{14.8cm}
\begin{center}
\vspace*{-0.2cm}
\psfrag{dsigpt}{\begin{Large}\hspace*{-0.50cm}
$\frac1{\sigma_{\rm SM}}\frac{d\sigma}{d p_{\rm T}}$\end{Large}[GeV$^{-1}$]}
\psfrag{sm}{\begin{small}Standard model\end{small}}
\psfrag{dvm}{\begin{small}(a)$\:d_V=-0.01,\:d_A=0,\:d_R=0$\end{small}}
\psfrag{dv}{\begin{small}(b)$\:d_V=0.01,\:d_A=0,\:d_R=0$\end{small}}
\psfrag{da}{\begin{small}(c)$\:d_V=0,\:d_A=0.05,\:d_R=0$\end{small}}
\psfrag{dr}{\begin{small}(d)$\:d_V=0,\:d_A=0,\:d_R=0.01$\end{small}}
\psfrag{dvar}{\begin{small}(e)$\:d_V=0.03,\:d_A=0.10,\:d_R=0.01$\end{small}}
\psfrag{pt}{\begin{large}\hspace*{-0.2cm}$p_{\rm T} $\end{large}}
\psfrag{gev}{\hspace*{-0.2cm}[GeV]}
\hspace*{-3.2cm}
\includegraphics[width=13cm, origin=c, angle=-90]{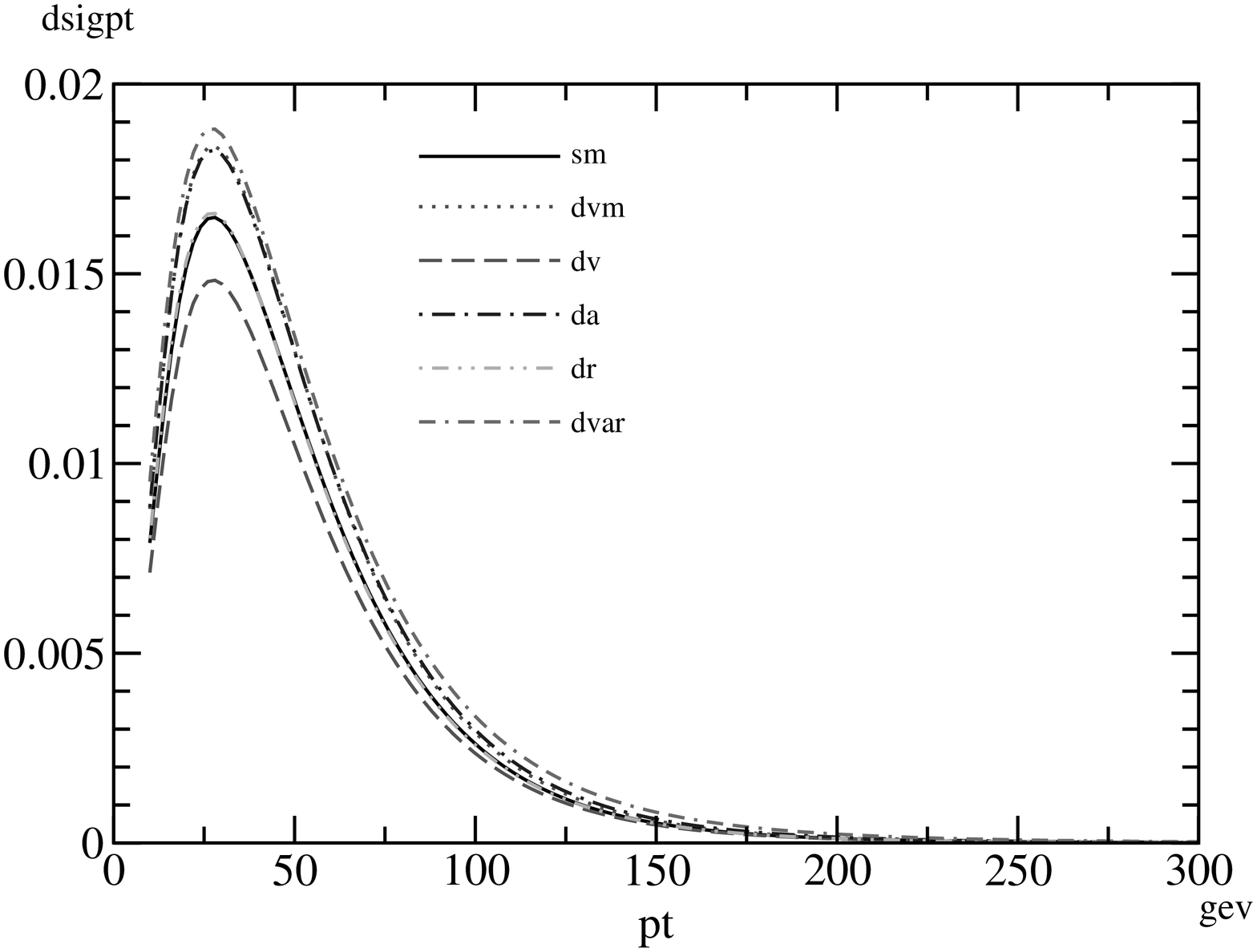}
\vspace*{-2.8cm}
\caption{The final-lepton transverse-momentum distribution normalized
by $\sigma_{\rm SM}$: LHC energy $\sqrt{s}=7$ TeV
\vspace{0.5cm}
}\label{lhcpt7}.
%\end{center}
%\end{minipage}
%\end{figure}
%%%%%%%%%%%%%%%%%%%%%%%%%%%%%%%%%%%%%%%%%%%%%%%%%%%%%%%%%%%%%%%%%%%%%%%%
%
%
%%%%%%%%%%%%%%%%%%%%%%%%%%%%%%%%%%%%%%%%%%%%%%%%%%%%%%%%%%%%%%%%%%
%
%   lhcpt14.eps
%
%%%%%%%%%%%%%%%%%%%%%%%%%%%%%%%%%%%%%%%%%%%%%%%%%%%%%%%%%%%%%%%%%%%%%
%\begin{figure}[H]
%\begin{minipage}{14.8cm}
%\begin{center}
\psfrag{dsigpt}{\begin{Large}\hspace*{-0.50cm}
$\frac1{\sigma_{\rm SM}}\frac{d\sigma}{d p_{\rm T}}$\end{Large}[GeV$^{-1}$]}
\psfrag{sm}{\begin{small}Standard model\end{small}}
\psfrag{dvm}{\begin{small}(a)$\:d_V=-0.01,\:d_A=0,\:d_R=0$\end{small}}
\psfrag{dv}{\begin{small}(b)$\:d_V=0.01,\:d_A=0,\:d_R=0$\end{small}}
\psfrag{da}{\begin{small}(c)$\:d_V=0,\:d_A=0.05,\:d_R=0$\end{small}}
\psfrag{dr}{\begin{small}(d)$\:d_V=0,\:d_A=0,\:d_R=0.01$\end{small}}
\psfrag{dvar}{\begin{small}(e)$\:d_V=0.03,\:d_A=0.10,\:d_R=0.01$\end{small}}
\psfrag{pt}{\begin{large}\hspace*{-0.2cm}$p_{\rm T} $\end{large}}
\psfrag{gev}{\hspace*{-0.2cm}[GeV]}
\hspace*{-3cm} % 
\includegraphics[width=13cm, origin=c, angle=-90]{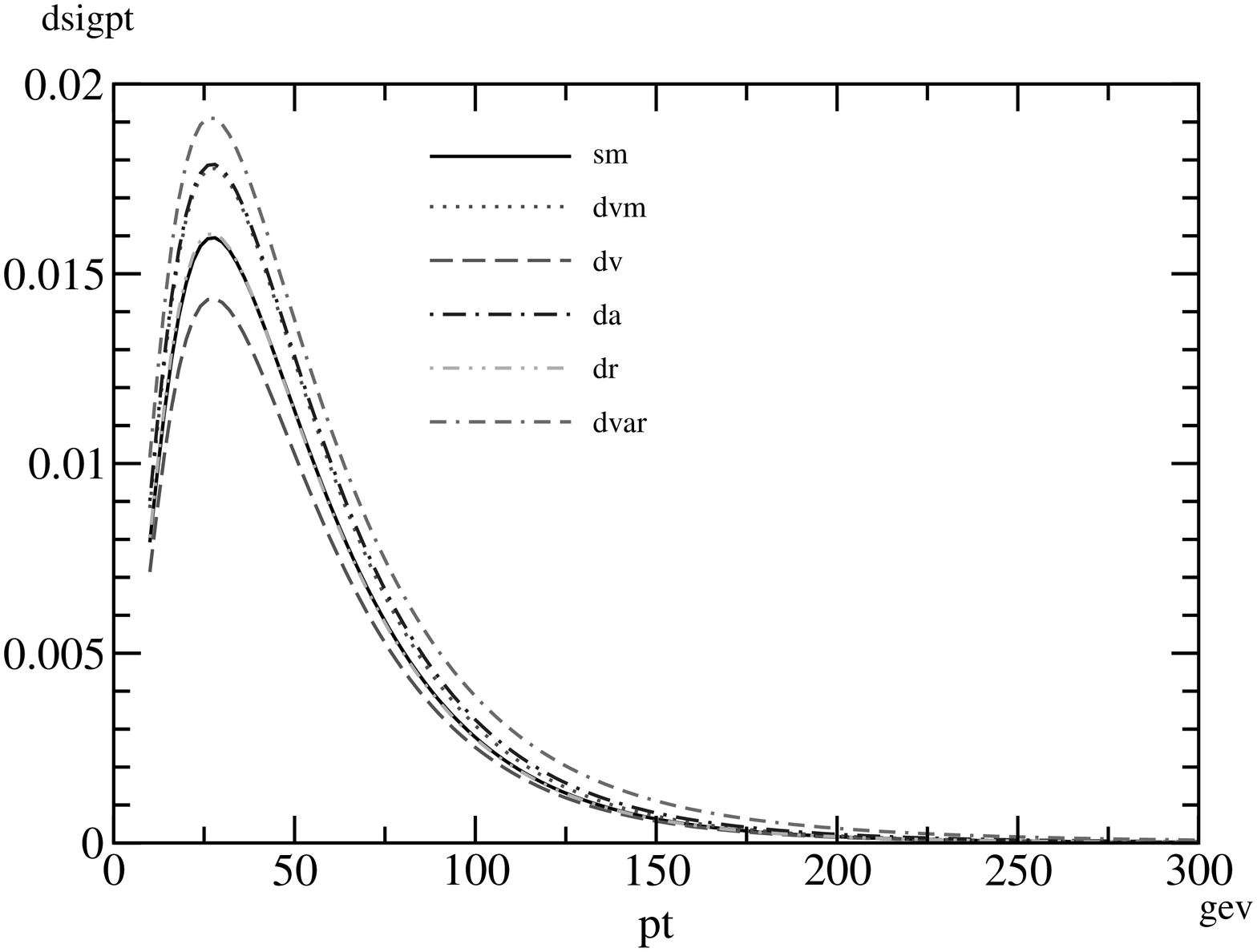}
\vspace*{-2.8cm}
\caption{The final-lepton transverse-momentum distribution normalized
by $\sigma_{\rm SM}$: LHC energy $\sqrt{s}=14$ TeV}\label{lhcpt14}.
\end{center}
\end{minipage}
\end{figure}
%%%%%%%%%%%%%%%%%%%%%%%%%%%%%%%%%%%%%%%%%%%%%%%%%%%%%%%%%%%%%%%%%%%%%

\newpage % \vskip 0.8cm

\noindent
the distributions are shown in Fig.\ref{tevpt}--Fig.\ref{lhcpt14}.

In these figures, the shapes of the curves are similar to those in Fig.\ref{tevel}--Fig.\ref{lhcel14}.
However, the magnitude of these $p_{\rm T}$ distributions
is  roughly two times larger than that of the $E_{\ell}$ distributions around their peak points.
The size of the deviations in percentage at $p_{\rm T}=50$ GeV is as follows:
\\
\underline{Tevatron}
\begin{equation}
\begin{array}{lll}
\delta({\rm a}) = +9.48\,\%,\phantom{0}~~~~~~& \delta({\rm b}) = -8.94\,\%,~~~~~~& \delta({\rm c}) = +2.14\,\%, \phantom{0}\\
\delta({\rm d}) = -0.49\,\%,      & \delta({\rm e}) = -17.42\,\%.      &                         \\
\end{array}
\end{equation}
\underline{LHC} (7 TeV)
\begin{equation}
\begin{array}{lll}
\delta({\rm a}) = +11.34\,\%,~~~~~~& \delta({\rm b}) = -10.03\,\%,~~~~~~& \delta({\rm c}) = +11.08\,\%, \\
\delta({\rm d}) = -0.39\,\%,       & \delta({\rm e}) = +14.60\,\%.      &                         \\
\end{array}
\end{equation}
%
%\noindent
\underline{LHC} (14 TeV)
\begin{equation}
\begin{array}{lll}
\delta({\rm a}) = +11.45\,\%,~~~~~~& \delta({\rm b}) = -10.02\,\%,~~~~~~& \delta({\rm c}) = +12.63\,\%, \\
\delta({\rm d}) = -0.34\,\%,       & \delta({\rm e}) = +20.97\,\%.      &                         \\
\end{array}
\end{equation}

% 555555555555555555555555555555555555555555555555555555555555
\sec{Summary and remarks}

We have studied possible anomalous $t\bar{t}g$-, $t\bar{t}gg$- and $tbW$-interaction effects
in the final-lepton distributions of $p\bar{p}/pp \to t\bar{t}X \to \ell^+ X'$
at Tevatron and LHC by assuming that there exists a new physics characterized by an
energy scale ${\mit\Lambda}$ and we only have standard-model particles/fields below
${\mit\Lambda}$. Under this assumption, all leading anomalous interactions are given 
by dimension-6 effective operators \cite{Buchmuller:1985jz,AguilarSaavedra:2008zc,Grzadkowski:2010es}. 
Based on the interaction Lagrangians composed of relevant effective
operators, we have derived analytical formulas of the parton-level cross sections 
of the processes $q\bar{q}/gg \to t\bar{t} X \to \ell^+ X'$ {\it for the first time
including both anomalous $C\!P$-conserving and $C\!P$-violating top-gluon couplings
as well as anomalous $tbW$ couplings at the same time}.\footnote{A similar work
    was done for the same processes (but in a different formalism) in \cite{Antipin:2008zx},
    where anomalous $C\!P$-violating top-gluon couplings and anomalous $tbW$ couplings were taken
    into account.}\
We then performed numerical calculations for the hadron-level processes at Tevatron and LHC
experiments. The results were shown in Fig.\ref{tevcl}--Fig.\ref{lhcpt14}, and then we came
to the following conclusions:
\begin{itemize}
 \item In case of $d_V \neq 0$ and $d_A \simeq 0$, we could observe discrepancy between
 the SM prediction and those with nonstandard effects. Moreover, comparing shapes depicted using
 parameter sets (a) and (b) in all figures, we saw that the opposite sign of $d_V$ could 
 induce opposite deviation from the SM predictions. This is simply because the leading $d_V$
 contribution comes from its linear terms, which fact will however be useful for determining
 the sign of this parameter.
 \item If both $d_V$ and $d_A$ are not so small, some nonstandard effects are expected to
 be observed. Furthermore, they could appear as different corrections at Tevatron and LHC:
 Look at Fig.\ref{tevel} and Fig.\ref{lhcel7}, for example. We see that deviations induced by
 parameter set (e) for Tevatron and LHC are in opposite direction from the SM prediction
 each other. The same holds true for other figures shown here. Those different deviations originate
 from the difference in the $t\bar{t}$-production mechanisms at Tevatron and LHC, i.e.,
 $q\bar{q}$-annihilation processes are dominant at Tevatron, while gluon-fusion processes
 dominate at LHC. This shows that Tevatron and LHC work complementarily to each other.
 \item In contrast to those $d_{V,A}$ contributions, that from $d_R$ was found to produce no sizable
 effects in the distributions we calculated here, and therefore it is difficult to measure its
 contributions in the processes on which we focused.
\end{itemize}

Finally, let us close this section with a couple of remarks.
First, we have limited our anomalous-coupling values to the inside of
the allowed region given in the appendix (Fig.\ref{allowed}). However this constraint
reflects $1\sigma$ level uncertainties. That is, their true values
might be outside that region, which leads to the possibility that larger deviations
from the SM prediction could be observed.
Secondly, we expressed all the anomalous interactions in terms of
several constant parameters. This is justified only when
$\sqrt{\hat{s}}\ll \mit{\Lambda}$ holds, which assumption might become
less accurate with increasing center-of-mass energy of LHC if the new
physics is just around the corner. In that case, unexpected nonstandard
effects could be measured.
Therefore, even if we might not discover any new particles at LHC,
it must be meaningful and important to increase its energy
in order to get signals from new physics beyond the standard model.
At any rate, we believe what was presented here will be one of the most
promising approaches to new physics at Tevatron and the current energy scale of
LHC ($\sqrt{s}=7$ TeV).

\vspace{0.7cm}
% AAAAAAAAAAAAAAAAAAAAAAAAAAAAAAAAAAAAAAAAAAAAAAAAAAAAAAA
\centerline{ACKNOWLEDGMENTS}

\vspace{0.3cm}
This work originates in part the doctor thesis of K.O. on $C\!P$ violation in
$p\bar{p} \to t\bar{t}$ and an encouraging comment to it by Bohdan Grzadkowski about
taking into account the final lepton, which comment we appreciate very much. This is
partly supported by the Grant-in-Aid for Scientific Research No.22540284 from
the Japan Society for the Promotion of Science. The algebraic calculations using
FORM were carried out on the computer system at Yukawa Institute for Theoretical
Physics (YITP), Kyoto University.

% \newpage % 
\vskip 0.8cm
\centerline{APPENDIX}

\vspace*{0.3cm}
% A1A1A1A1A1A1A1A1A1A1A1A1A1A1A1A1A1A1A1A1A1A1A1A1A1A1A1A1A1
% \noindent \hskip -0.46cm % \hskip -0.72cm
\noindent
{\bf Constraints on the anomalous top-gluon couplings}
% {\bf A. Constraints on the anomalous top-gluon couplings}

\noindent
In our previous analysis on $d_V$ and $d_A$ through the total cross section of $t\bar{t}$ productions
\cite{Hioki:2009hm} we pointed out that LHC data could give a stronger constraint
on them, which would be hard to obtain from Tevatron data alone. We then showed in
\cite{Hioki:2010zu} that the first CMS measurement of this quantity \cite{Khachatryan:2010ez}
actually made it
possible. That is, we have obtained a stronger constraint on $d_{V,A}$ by combining
the CDF/D0 data \cite{Teva-data}
\begin{eqnarray}
&&\!\!\!\!\!\!\!\!\!
\sigma_{\rm exp}
= 7.02 \pm 0.63\ {\rm pb}\: \ \ ({\rm CDF}:\:m_t=175\:{\rm GeV}) \\
&&\!\!\!\!\!\!\!\!\!
\phantom{\sigma_{\rm exp}}
= 8.18^{\ +\ 0.98}_{\ -\ 0.87}\ {\rm pb}\ \ \ \ ({\rm D0}:\:m_t=170\:{\rm GeV})
\end{eqnarray}
with the CMS data
\begin{equation}
\sigma_{\rm exp}
= 194 \pm 72\,({\rm stat.}) \pm 24\,({\rm syst.}) \pm 21\,({\rm lumi.}) \ {\rm pb}\: \ \ (m_t=172.5\:{\rm GeV})
\end{equation}
than in the analysis with the above CDF/D0 data alone. Since we now also have ATLAS data \cite{Aad:2010ey}
\begin{equation}
\sigma_{\rm exp}
= 145 \pm 31 ^{+42}_{-27}\ {\rm pb}\: \ \ (m_t=172.5\:{\rm GeV}),
\end{equation}
where the first uncertainty is statistical and the second systematic, it is worth carrying out the same
analysis again with all the data available here.\footnote{We do not repeat describing the detail of
    the calculations here and leave it to \cite{Hioki:2009hm}.}

In this analysis, we need the absolute value of the cross section, for which we cannot neglect the
QCD radiative corrections. As for such corrected SM contribution, we took the Next-to-Leading-Order (NLO)
cross section
\begin{equation}
\sigma_{\rm SM}^{\rm NLO} = 157.5^{+23.2}_{-24.4}\ {\rm pb}
\end{equation}
in \cite{Hioki:2010zu}, which was used by the CMS \cite{Khachatryan:2010ez}.
We here, however, take account of the NNLO value
\begin{equation}
\sigma_{\rm SM}^{\rm NNLO}
= 164.6^{+11.4}_{-15.7}\ {\rm pb}\: \ \ (m_t=172.5\:{\rm GeV})
\end{equation}
as in \cite{Aad:2010ey}.

\vspace*{0.4cm}
The result is shown in Fig.\ref{allowed}, where the shaded part is the $d_{V,A}$ region allowed
by Tevatron and LHC data altogether. There does not seem to be any big difference from the result
in \cite{Hioki:2010zu}, but the allowed area has become a bit narrower by adding the ATLAS data.
All the parameter values used in the main text were taken from inside this region.

\baselineskip=20pt plus 0.1pt minus 0.1pt

 \vspace*{0.8cm}
% RRRRRRRRRRRRRRRRRRRRRRRRRRRRRRRRRRR

\end{document}